\documentclass[11pt]{article}
\usepackage[a4paper, margin = 1in]{geometry}
\usepackage[numbers]{natbib}
\usepackage[titletoc,page]{appendix}

\usepackage{graphicx}
\usepackage{setspace}
\usepackage{gensymb}
\usepackage{enumerate}
\usepackage{kotex}
\usepackage{abstract,lipsum}
\usepackage[colorlinks=true,linkcolor=blue,citecolor=blue,urlcolor=blue]{hyperref}

\usepackage{varioref}

\usepackage{titlesec}
\titleformat{\section}[block]{\normalfont\Large\bfseries}{\thesection}{1.0em}{}
\titleformat{\subsection}[block]{\normalfont\Large\bfseries}{\thesubsection}{1.0em}{}

\usepackage{amsmath}
\usepackage{amssymb}
\usepackage{amsthm}
\usepackage{amsfonts}
\usepackage{upgreek}
\usepackage{mathtools}
\usepackage{mathrsfs}
\usepackage{bm}
\newcommand\defeq{\stackrel{\mathclap{\normalfont\tiny\mbox{def}}}{=}}
\usepackage[symbol]{footmisc}

\usepackage{color}
\usepackage{multirow}
\usepackage{algorithm,algorithmic}
\usepackage{verbatim}

\usepackage{tcolorbox}
\usepackage[english]{babel}
\theoremstyle{remark}

\begin{document}

\begin{center}
       \fontsize{15pt}{15pt}\selectfont \textbf{Stiffness, strength, energy dissipation and reusability in heterogeneous architected polycrystals}
       
       \vspace*{0.3in}
       \fontsize{9.5pt}{9.5pt}\selectfont Seunghwan Lee$^{1}$, Hansohl Cho$^{1,2\dagger}$,  \\
       \vspace*{0.3in}
       \fontsize{9pt}{9pt}\selectfont $^{1}$Department of Aerospace Engineering, $^{2}$Department of Mechanical Engineering, Korea Advanced Institute of Science and Technology, Daejeon, 34141, Republic of Korea \\ 

\vspace*{0.2in}
\fontsize{9.5pt}{9.5pt}\selectfont E-mail: $^\dagger$hansohl@kaist.ac.kr
\end{center}

\renewenvironment{abstract}
{\small 
\noindent \rule{\linewidth}{.5pt}\par{\noindent \bfseries \abstractname.}}
{\medskip\noindent \rule{\linewidth}{.5pt}
}

\vspace*{0.3in}
\onehalfspacing
\begin{abstract}
\fontsize{10pt}{10pt}\selectfont
We design, fabricate and test heterogeneous architected polycrystals, composed of hard plastomers and soft elastomers, which thus show outstanding mechanical resilience and energy dissipation simultaneously. Grain boundaries that separate randomly oriented single crystalline grains is carefully designed, first enabling coherent connectivity and strength in the grain boundary regions throughout the polycrystalline network. By combining experiments and numerical simulations on 3D-printed prototypes, we show that the interplay between grain interiors and grain boundaries is responsible for the grain-size effects emerging in these architected materials, analogous to those in their atomic or metallic counterparts. Furthermore, direct visualization of inter- and intra-grain deformation and failure mechanisms at the macroscopic scale reveals that crystallographic texture throughout the polycrystalline aggregates plays a fundamental role in the key mechanical features in our new heterogeneous polycrystals. Our results show that the engineered grain boundary and crystallographic texture not only modify the highly resilient yet dissipative global responses but also critically influence reusability in this new class of architected materials.
\\
\end{abstract} 

\doublespacing
\section{Introduction}
Geometric and topological structuring of constituent materials leads to enhanced mechanical performance and functionality which cannot be attained through compositions in heterogeneous materials. Furthermore, recent fabrication breakthroughs have enabled the creation of heterogeneous “architected” materials exhibiting simultaneous resilience and energy dissipation, which are known to be thermodynamically competing, by using intricate structuring and ordering of subdomains. New heterogeneous architected materials with extraordinary resilience and energy dissipation have been demonstrated, mainly inspired by the geometric and topological features in naturally formed materials and composites over the past decade \cite{ortiz2008bioinspired,bruet2008materials,wegst2015bioinspired,ritchie2021toughening,fernandes2021mechanically}. More recently, unique microstructures often found in man-made heterogeneous materials have also provided material scientists and designers ample opportunities to realize augmented mechanical functionalities. Examples include heterogeneous soft materials with tunable elastic and inelastic features inspired by random microstructures in block copolymers and phase-separated polymers \cite{wang2011co,lee2012periodic,cho2013dissipation,vidyasagar2018microstructural,siddique2021nanonetwork,lee2023polyurethane,fernandez2024elastic,cho2024large}.

In addition to the complex geometries formed in natural or synthetic composites and block copolymers, single- and polycrystalline subdomains ubiquitous in metallic materials are also receiving great attention for the rational design of new heterogeneous (or cellular) architected materials with unprecedented mechanical properties \cite{cho2016engineering,tancogne2016additively,liontas20173d,pham2019damage,liu2021origin,li2022effects,yu2023design,liu2024spatially,tian2024programmable}. Not only are macro- and mesoscopic architected crystals found to exhibit interesting mechanical properties, analogous to their atomic counterparts, but also they have allowed us to better address the metallurgical mechanisms associated with dislocation- or grain boundary-mediated plastic deformations and failure processes by visualizing their structures and dynamics directly at both single- \cite{liontas20173d,schall2004visualization,schall2006visualizing,kim2024work} and polycrystal levels \cite{pham2019damage,liu2021origin,tian2024programmable}. Further, by introducing simple boundaries between single crystalline grains, polycrystal-inspired architected materials have been able to attain extraordinary damage tolerance not reported in previous studies of mechanical meta-materials built on single crystal lattices \cite{pham2019damage,liu2021origin,li2022effects,yu2023design}. It should also be noted that most of these architected polycrystals have made use of simple twin boundaries that separate their single crystalline subdomains. Albeit simple, such twinned architectures offer ample opportunities successfully to translate some polycrystalline metallurgical principles into the designs of high strength and damage-tolerant architected materials. However, polycrystal-like architected materials with simple twin boundaries lack the realistic complexities widely found at the grain boundaries that separate randomly oriented single crystalline subdomains. Indeed, the complex geometric and topological features at the grain boundaries have rendered the rational design and demonstration of polycrystal-inspired materials more challenging.

In this work, we have designed and demonstrated architected polycrystalline materials using a combination of experiments and numerical simulations. The architected polycrystals are composed of plastomeric “hard” components and elastomeric “soft” components and thus offer both mechanical resilience and energy dissipation capabilities. Towards this end, we fabricated and tested prototypes of architected polycrystals using a high-precision, multi-material three-dimensional printer that uses thermoplastic polymers and rubbery polymers for the hard and soft domains, respectively. Importantly, the grain boundary network that separates randomly oriented single crystalline subdomains is carefully designed, enabling coherent connectivity and strength throughout the grain boundary regions in the 3D-printed prototypes. Through the careful design of polycrystalline architectures with coherent grain boundaries, we quantitatively explore the grain-size effects on elastic stiffness and plastic strength in the architected polycrystals, by which an interplay between grain interiors and grain boundaries for the size-dependent elastic and plastic features is addressed in detail. More importantly, the combined results of experiments and numerical simulations demonstrate that tailoring crystallographic texture throughout the polycrystalline aggregates not only influences the grain size-dependent stiffness and strength significantly but also plays a crucial role in the resilience, load transfer, energy dissipation, and failure mechanisms of these architected polycrystals when subjected to multiple cyclic loading and unloading conditions.

The paper is organized as follows. A design procedure to construct the heterogeneous architected polycrystals is presented in Section \ref{Designing heterogeneous architected polycrystals} and Appendix \ref{Microstructural details in heterogeneous architected polycrystals}. We present the emergence of grain-size dependent elastic and inelastic features in these architected polycrystals through both experiments and numerical simulations in Section \ref{Grain-size effect on elastic stiffness and inelastic strength}. Then, we demonstrate energy dissipation and load transfer capabilities in our heterogeneous polycrystalline materials weakly or strongly textured upon multiple cyclic loading, unloading and reloading conditions in Section \ref{Energy dissipation, local failure and reusability}, by which we elucidate the role of grain boundaries and textures in reusability of this new class of architected materials. We close with concluding remarks on our findings and the main contributions of this work in Section \ref{Conclusion}. Experimental and micromechanical modeling procedures to characterize the stiffness, strength, energy dissipation and reusability are presented in Appendix \ref{Experimental and numerical procedures} together with detailed information on the large strain behavior of each of the constitutive materials used in the heterogeneous polycrystals in Appendix \ref{constitutive modeling}. We finally present further discussion on energy dissipation, local failure and load transfer mechanisms in our polycrystals in Appendix \ref{Further analysis on energy dissipation, local failure and reusability}.

\section{Designing heterogeneous architected polycrystals}
\label{Designing heterogeneous architected polycrystals}

\subsection{Grain interiors and grain boundaries}
\begin{figure}[b!]
    \centering
    \includegraphics[width=1.0\textwidth]{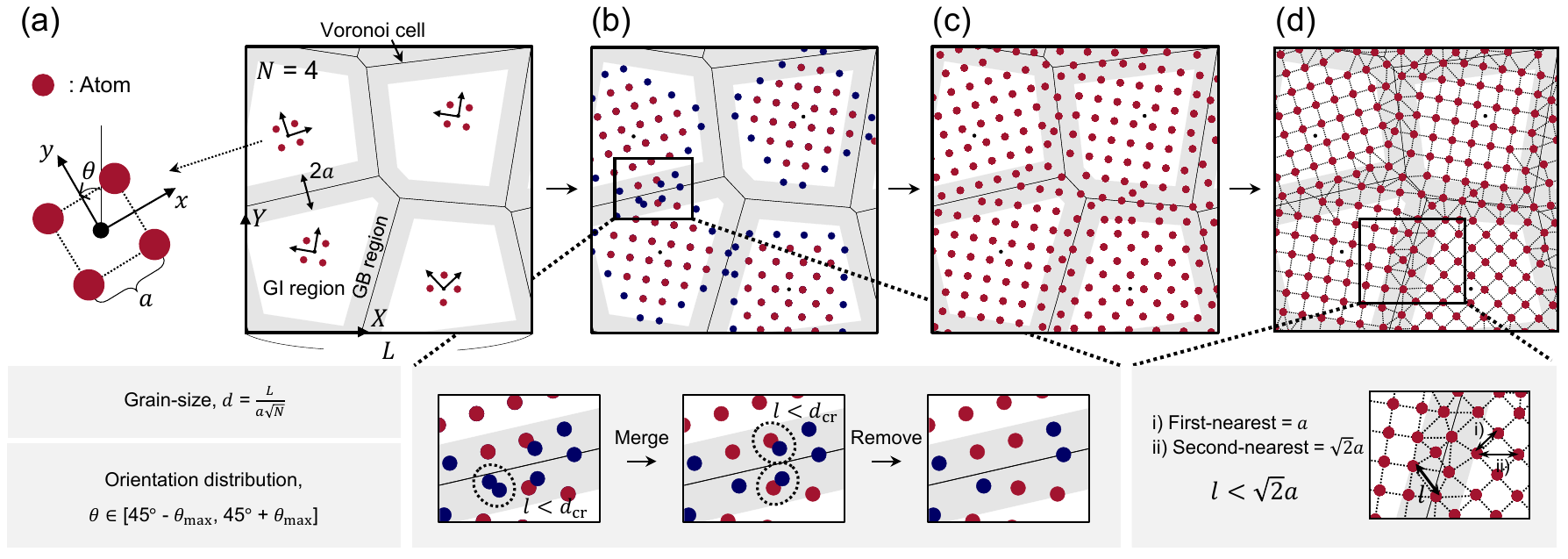}
    \caption{Design procedure of heterogeneous architected polycrystals. (a) Two-dimensional square unit-cell with lattice constant $a$ (far left inset) and $N$ Voronoi cells divided into grain interior and grain boundary regions; here, $N = 2 \times 2$ is presented for brevity. (b) Rules of merging and removing atoms in grain boundary regions. (c) A resulting arrangement of atoms and (d) atomic pairs for strut-connecting.}
    \label{fig:design}
\end{figure}
Figure \ref{fig:design} displays a facile yet systematic strategy for designing a representative volume element (RVE) of the heterogeneous architected polycrystals. As illustrated in Figure \ref{fig:design}a, we use a two-dimensional square unit-cell with lattice constant $a$ and crystal orientation $(x,y)$ rotated counterclockwise from the $Y$-axis of a sample coordinate system $(X,Y)$ by a crystal orientation angle $\theta \in [0^\circ,90^\circ]$. The distribution of the crystal orientations strongly influences texture of the architected polycrystals. In order to tailor the crystallographic texture in the polycrystals, the crystal orientation angle $\theta$ in each grain is selected from the range, $[45^\circ - \theta_{\mathrm{max}},45^\circ + \theta_{\mathrm{max}}]$, where $\theta_{\mathrm{max}}$ is set to $45^\circ$ (i.e., $\theta \in [0^\circ,90^\circ]$) for weakly textured architected polycrystals (see Figure \ref{fig:size_low}) and to $15^\circ$ (i.e., $\theta \in [30^\circ,60^\circ]$) for strong textures (see Figure \ref{fig:size_high}). Voronoi cells are generated from $N$ seed points randomly distributed in an $L \times L$ domain under periodic boundary conditions. The $N$ Voronoi cells, with an average area of $L^2/N$, are then divided into grain interior (GI) and grain boundary (GB) regions; here, the thickness of the GB regions is assumed to be $2a$ (see Figure \ref{fig:gb_design} in Appendix \ref{section:constructing} for details on the construction of the GB regions). Furthermore, the average grain size is defined as
$d=L/(a\sqrt{N})$, where $L/\sqrt{N}$ is the average Voronoi cell size. In other words, a decrease in the grain size leads to an architected polycrystal with fewer “square” unit-cells in each of the grains, resulting in a higher volume fraction of the GB region. In this work, the non-dimensional grain sizes are taken to be $d=3.75$ and $d=15$ to explore the grain size-dependent elastic and inelastic features in the architected polycrystals (see Figures \ref{fig:size_low} and \ref{fig:size_high}). For each seed point, we radially arrange the atoms at each step by mimicking the crystal growth, as displayed in Figure \ref{fig:design}b (see Movie S1 for details); for clarity, the atoms newly arranged in the intermediate step are highlighted in blue. In order to tailor “connectivity or strength” throughout the grain boundary networks, the number of atoms embedded in the GB regions is varied by introducing a critical distance, $d_{\mathrm{cr}}$, defined as the minimum distance between neighboring atoms. When the distance between the neighboring atoms (in blue) is shorter than $d_{\mathrm{cr}}$ (i.e., $l < d_{\mathrm{cr}}$), the atoms are merged into a new atom (in blue) located at the midpoint, as illustrated in the lower inset of Figure \ref{fig:design}b (see Figure \ref{fig:gb_strength} in Appendix \ref{section:tailoring} for more specific details on the merging procedure of the neighboring atoms). Then, all of the atoms (in blue) within the critical distance from the previously arranged atoms (in red) are removed. The resulting arrangement of atoms throughout the grain boundary network is displayed in Figure \ref{fig:design}c. The atomic pairs identified by Delaunay triangulation are then strut-connected only when their distance is shorter than $\sqrt{2}a$ (i.e., the distance between the second-nearest neighbors in the square unit-cell), as shown in Figure \ref{fig:design}d. This procedure enables the simultaneous construction of GI regions with square lattices and GB regions with tailored connectivity that varies depending on $d_{\mathrm{cr}}$. Because atoms are less removed or merged as $d_{\mathrm{cr}}$ decreases, more atoms are embedded in the GB regions, by which the GB network exhibits greater connectivity (or strength); in this work, the critical distances are taken to be $d_{\mathrm{cr}}=0.5a$ and $d_{\mathrm{cr}}=0.8a$ for high and low GB connectivities (or strengths), respectively.

\subsection{Sample fabrication}
Prototypes of the polycrystalline RVEs were fabricated using a high-resolution multi-material 3D printer (Connex3 objet260, Stratasys Inc.), as presented in Appendix \ref{section:prototype}. VeroWhitePlus$^{\mathrm{TM}}$ (a thermoplastic polymer) and TangoPlus$^{\mathrm{TM}}$ (a rubbery polymer) were used for the hard polycrystalline networks and the isolated soft domains, respectively. The sample dimensions are 120mm x 120mm (width x height) and 10mm (depth). Given that the resolution of the 3D printer is $\sim$ 30$\upmu$m, the minimum feature size of substructures (i.e., strut thickness) is set to be $\sim$ 300$\upmu$m in the samples.

We note that, as displayed in Figure \ref{fig:prototype} in Appendix \ref{section:prototype}, the architected polycrystals studied in this work are not cellular materials; rather, they are “two-phase” materials, where the “hard” polycrystalline networks (bright regions in the experimental images; e.g., see also Figures \ref{fig:size_low} and \ref{fig:size_high}) were fabricated using a thermoplastic polymer while the isolated “soft” domains (dark regions in the experimental images) were made of a rubbery polymer. The elastic modulus ratio of the hard to the soft components is $\sim$ 2000. Furthermore, the thermoplastic hard component exhibits highly energy dissipative elasto-plastomeric behavior with high yield as well as high flow stresses while the rubbery soft component is hyperelastic with negligible hysteresis upon unloading. See Appendix \ref{section:large} on the constitutive behavior of each of the two constituent materials. We also note that the volume fraction of the hard domain (i.e., polycrystalline networks) in the architected polycrystals is set to $\mathrm{v}_{\mathrm{hard}} = 40\%$.

\section{Grain-size effect on elastic stiffness and inelastic strength}
\label{Grain-size effect on elastic stiffness and inelastic strength}
\begin{figure}[t!]
    \centering
    \includegraphics[width=1.0\textwidth]{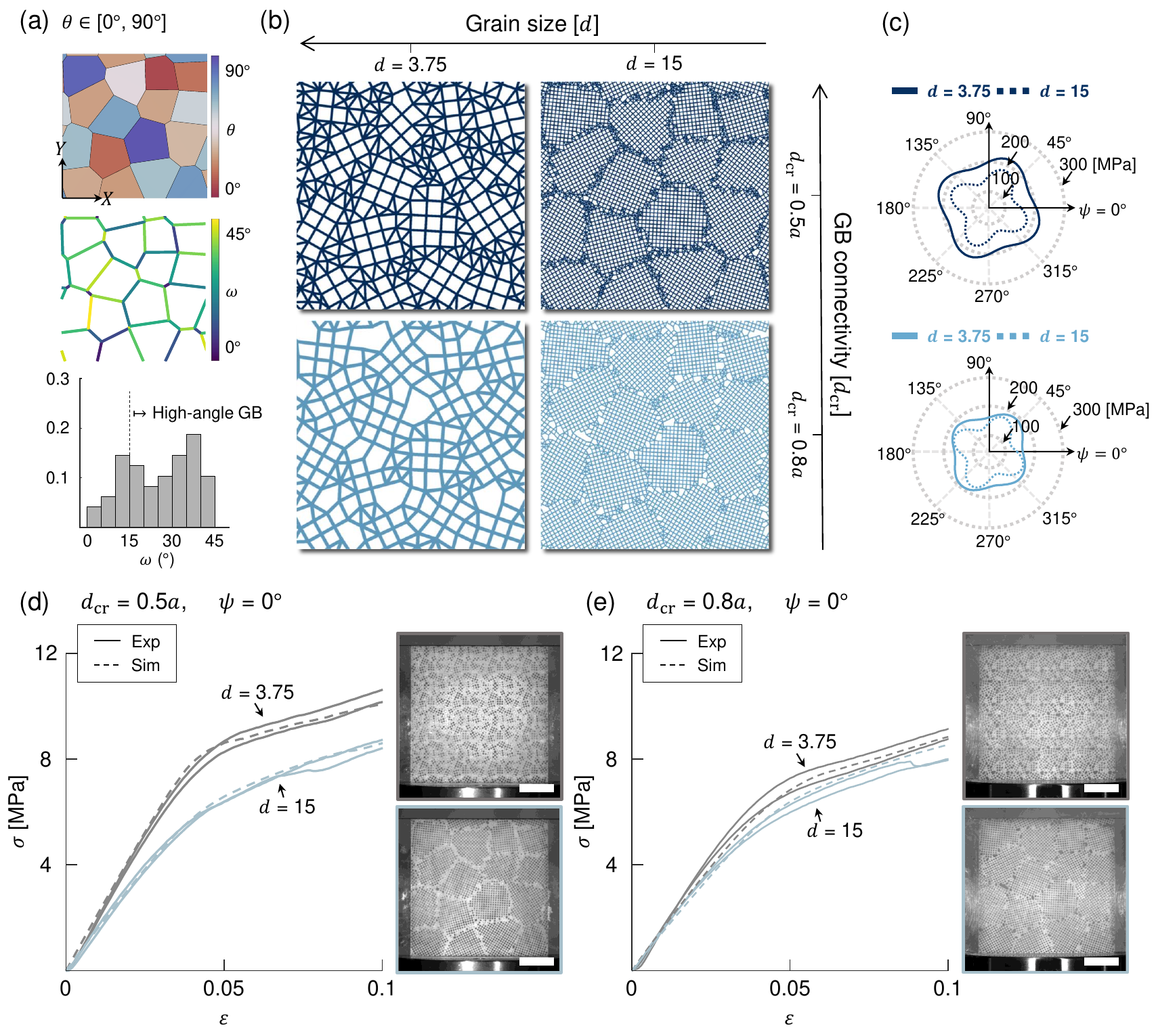}
    \caption{Grain-size effects on anisotropy, elastic stiffness and inelastic strength in weakly textured ($\theta_{\mathrm{max}}=45^\circ$) polycrystals. (a) Distributions of crystal orientation angle $\theta$ (in grain interiors) and disorientation angle $\omega$ (at grain boundaries), along with a histogram of $\omega$. (b) Polycrystalline RVEs with varying grain sizes ($d$) and GB connectivities ($d_{\mathrm{cr}}$); here, only hard domains are displayed. (c) Loading direction ($\psi$)-dependent elastic moduli in polycrystalline RVEs with $d_{\mathrm{cr}}=0.5a$ (upper inset) and $0.8a$ (lower inset); solid lines: $d=3.75$, dashed lines: $d=15$. Stress-strain curves in experiments (solid lines) and numerical simulations (dashed lines) at a strain rate of 0.01$\mathrm{s}^{-1}$ for polycrystalline RVEs with (d) $d_{\mathrm{cr}}=0.5a$ and (e) $0.8a$ loaded at $\psi=0^\circ$ (gray: $d=3.75$, blue: $d=15$) together with experimental images of undeformed configurations for visual aid. Scale bars: 3cm.}
    \label{fig:size_low}
\end{figure}
First, we investigated the grain size-dependent elastic and inelastic features in the heterogeneous architected polycrystals at small to large strains; detailed information on large strain mechanical testing of the 3D-printed samples can be found in Appendix \ref{section:mechanical}. Here, the number of grains in each of the heterogeneous architected polycrystals was set to $N=16$. The crystal orientation angle $\theta$ in each grain was randomly selected with a range of $[0^\circ, 90^\circ]$, for which $\theta_{\mathrm{max}}$ was set to $45^\circ$, as displayed in Figure \ref{fig:size_low}a. We then calculated the disorientation angle, $\omega$, defined as the minimum angle required to align two adjacent crystal lattices perfectly \cite{kocks2000texture,frary2004percolation,ratanaphan2015grain} in these heterogeneous architected polycrystals with $N=16$, as shown in the middle inset of Figure \ref{fig:size_low}a (see Figure \ref{fig:single_grain}a in Appendix \ref{section:microstructural} for details). Consequently, as shown in the lower inset of Figure \ref{fig:size_low}a of the distribution of the disorientation angles, the architected materials possess a high fraction ($\sim$ 75\%) of high-angle grain boundaries (or GBs) with $\omega>15^\circ$ \cite{frary2004percolation,hughes1997high,knezevic2014three,wang2016spectral}. We then designed RVEs of heterogeneous architected polycrystals with varying grain sizes ($d=3.75$ and $15$) and critical distances ($d_{\mathrm{cr}}=0.5a$ and $0.8a$), as illustrated in Figure \ref{fig:size_low}b. The polycrystalline RVEs with smaller grain sizes were found to have a higher volume fraction of GB regions. Furthermore, the degree of connectivity increases throughout the GB regions in the RVEs as $d_{\mathrm{cr}}$ decreases. We then conducted micromechanical analysis of the polycrystalline RVEs subjected to a periodic boundary condition and to macroscopic plane-strain compression. See Appendix \ref{section:micromechanical} on the micromechanical modeling procedure. Figure \ref{fig:size_low}c shows the loading direction-dependent elastic modulus in these RVEs; the loading direction is defined as the angle $\psi$ rotated clockwise from the $Y$-axis of the sample coordinate system (see Figure \ref{fig:single_grain}b in Appendix \ref{section:microstructural} for details on the angle of loading direction, $\psi$). For two different GB connectivities, with $d_{\mathrm{cr}}=0.5a$ and $d_{\mathrm{cr}}=0.8a$, the RVEs are shown to exhibit more isotropic elastic responses throughout the loading directions as the size of the grain decreases, attributed to the higher volume fraction of the amorphous GB regions in the RVEs with $d=3.75$; the degrees of anisotropy $A$, defined by the ratio of the minimum elastic modulus to the maximum elastic modulus throughout the loading directions, are found to be $A=0.83$ and $0.71$ for $d=3.75$ and $15$, respectively ($d_{\mathrm{cr}}=0.5a$); and $A=0.85$ and $0.72$ for $d=3.75$ and $15$, respectively ($d_{\mathrm{cr}}=0.8a$). Furthermore, the elastic stiffnesses are found to be greater in RVEs with a smaller grain size ($d=3.75$) throughout all loading directions ($\psi$). Importantly, the effect of the grain size on the elastic responses is more apparent in RVEs with greater GB connectivity ($d_{\mathrm{cr}}=0.5a$).

We also show that GB connectivity significantly influences the grain-size effect on the inelastic strength beyond the initial elastic regime in these polycrystalline materials in both experiments and numerical simulations. Figures \ref{fig:size_low}d and \ref{fig:size_low}e show the grain size ($d=3.75$ and $d=15$)-dependent stress-strain curves in the RVEs with $d_{\mathrm{cr}}=0.5a$ and $d_{\mathrm{cr}}=0.8a$; here, the loading direction was taken to be $\psi=0^\circ$. In addition to the grain size-dependent elastic responses, the inelastic strength is shown to be greater in the RVEs with smaller grains (gray lines in Figures \ref{fig:size_low}d and \ref{fig:size_low}e); further, the grain-size effect on the inelastic stress-rollover (or yield) is more apparent in the RVEs with greater GB connectivity ($d_{\mathrm{cr}}=0.5a$). The grain-size effects on the elastic stiffness and inelastic strength are further underpinned by the microstructural details in these polycrystalline materials. We examined the average nodal connectivity $\mathrm{Z}$, especially in the GB regions; the nodal connectivity, also known as the coordination number, has been widely used to quantify the “bond” stiffness and strength in diverse architected materials \cite{fleck2010micro,gross2019correlation,luan2023data,magrini2024control,lee2024extreme}. The average nodal connectivity of the “GB regions”, denoted $\mathrm{Z}_{\mathrm{GB}}$, for each of the RVEs are found to be $\mathrm{Z}_{\mathrm{GB}}=5.35$ and $5.11$ for $d=3.75$ and $15$ ($d_{\mathrm{cr}}=0.5a$); $\mathrm{Z}_{\mathrm{GB}}=4.46$ and $4.35$ for $d=3.75$ and $15$ ($d_{\mathrm{cr}}=0.8a$); we note that the nodal connectivity of the crystal lattices within the grain interiors is 4. More importantly, as the grain size is decreased ($d=15$ to $d=3.75$), the average nodal connectivity $\mathrm{Z}$ for the polycrystalline RVEs including both GI and GB regions is found to increase by $\sim$ 20\% with $d_{\mathrm{cr}}=0.5a$ (i.e., $\mathrm{Z}=4.4$ to $\mathrm{Z}=5.24$). However, it only increases by $\sim$ 7\% in the RVEs with $d_{\mathrm{cr}}=0.8a$ (i.e., $\mathrm{Z}=4.10$ to $\mathrm{Z}=4.37$). In other words, the higher fraction of GB regions with greater connectivity (or strength) in the RVEs with smaller grains more effectively enhances both elastic stiffness and inelastic strength, giving rise to the more apparent grain-size effects in these architected polycrystals.

\begin{figure}[t!]
    \centering
    \includegraphics[width=0.95\textwidth]{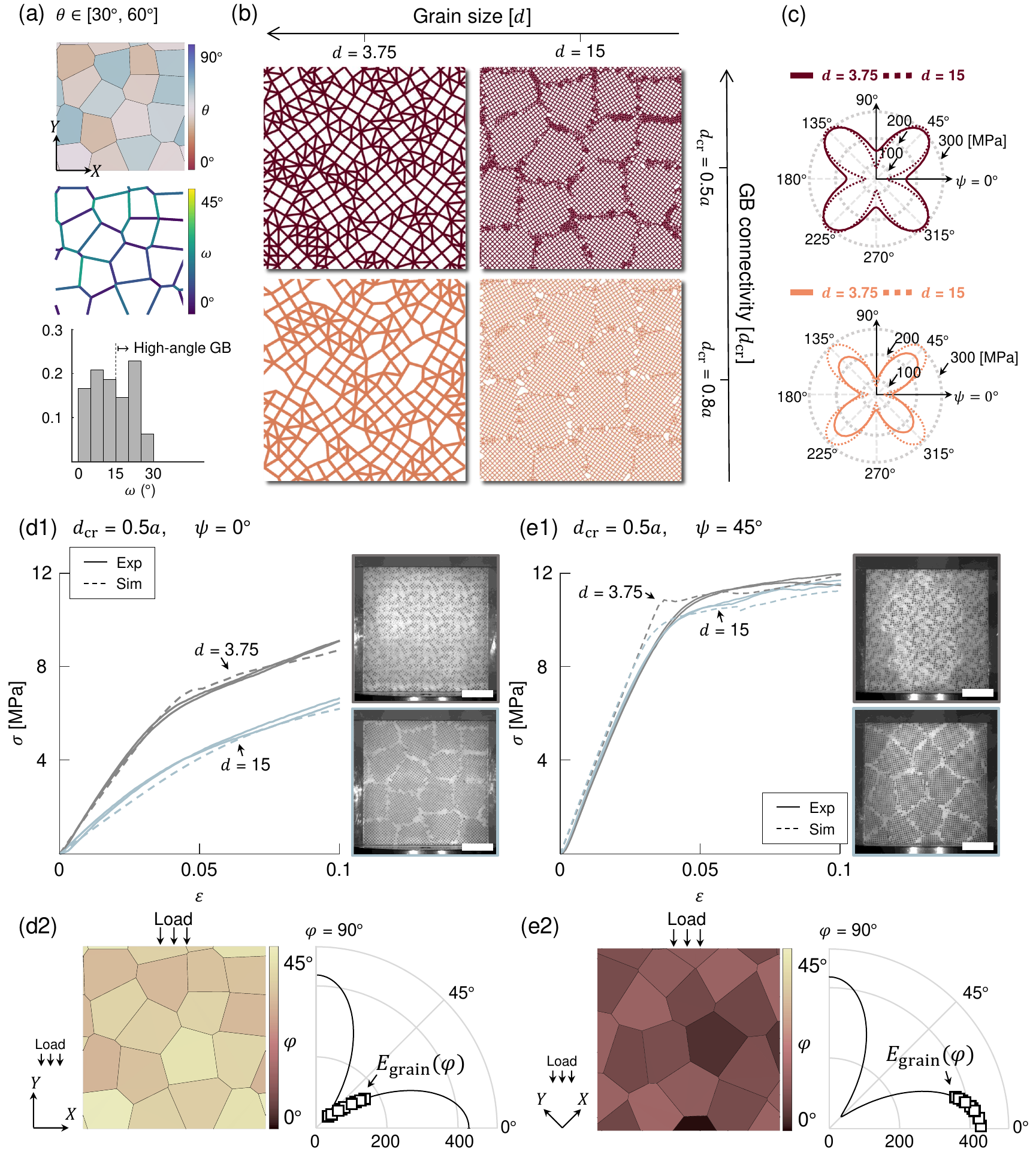}
    \caption{Grain-size effects on anisotropy, elastic stiffness and inelastic strength in strongly textured ($\theta_{\mathrm{max}}=15^\circ$) polycrystals. (a) Distributions of crystal orientation angle $\theta$ (in grain interiors) and disorientation angle $\omega$ (at grain boundaries), along with a histogram of $\omega$. (b) Polycrystalline RVEs with varying grain sizes ($d$) and GB connectivities ($d_{\mathrm{cr}}$); here, only hard domains are displayed. (c) Loading direction ($\psi$)-dependent elastic moduli in polycrystalline RVEs with $d_{\mathrm{cr}}=0.5a$ (upper inset) and $0.8a$ (lower inset); solid lines: $d=3.75$, dashed lines: $d=15$. Stress-strain curves in experiments (solid lines) and numerical simulations (dashed lines) at a strain rate of 0.01$\mathrm{s}^{-1}$ for polycrystalline RVEs with $d_{\mathrm{cr}}=0.5a$ loaded at (d1) $\psi=0^\circ$ and (e1) $45^\circ$ (gray: $d=3.75$, blue: $d=15$) together with experimental images of undeformed configurations for visual aid. Distributions of $\varphi$ and the corresponding directional elastic modulus in each grain (i.e., $E_{\mathrm{grain}} (\varphi)$) when loaded at (d2) $\psi=0^\circ$ and (e2) $\psi=45^\circ$. Note that most of the grains are aligned with $\varphi=45^\circ$ when loaded at $\psi=0^\circ$ ($\varphi=0^\circ$ when loaded at $\psi=45^\circ$) in these strongly textured polycrystals. Scale bars: 3cm.}
    \label{fig:size_high}
\end{figure}

The grain size-dependent elastic and inelastic features are further presented in Figure \ref{fig:size_high} on the heterogeneous architected polycrystals with a strong texture. Here, $\theta_{\mathrm{max}}$ was set to $15^\circ$; i.e., $\theta$ was randomly selected within the range of $[30^\circ, 60^\circ]$, leading to a relatively high fraction of low-angle GBs ($\sim$ 56\%; $\omega<15^\circ$), as shown in the histogram in Figure \ref{fig:size_high}a. We then constructed polycrystalline RVEs, as displayed in Figure \ref{fig:size_high}b, with the same grain sizes ($d=3.75$ and $15$) and critical distances ($d_{\mathrm{cr}}=0.5a$ and $0.8a$) used in the weakly textured RVEs with $\theta_{\mathrm{max}}=45^\circ$ presented in Figure \ref{fig:size_low}. A micromechanical analysis showed significant elastic anisotropy in all of the RVEs with $\theta_{\mathrm{max}}=15^\circ$ due to the strong texture, as presented in Figure \ref{fig:size_high}c of the directional elastic stiffness in the RVEs. Furthermore, as the grain size became smaller, the elastic anisotropy was found to lessen, with $A=0.54$ and $0.28$ for $d=3.75$ and $15$, respectively, with $d_{\mathrm{cr}}=0.5a$; and $A=0.42$ and $0.26$ for $d=3.75$ and $15$ with $d_{\mathrm{cr}}=0.8a$. Interestingly, the effect of the grain size on elastic stiffness and inelastic strength in these heterogeneous polycrystals strongly textured with $\theta_{\mathrm{max}}=15^\circ$ is shown to be highly dependent on the loading direction ($\psi$). As presented in Figure \ref{fig:size_high}d1 of the measured and numerically simulated stress-strain curves in the RVEs loaded at $\psi=0^\circ$, as the grain size is decreased, elastic stiffness and inelastic strength both increase significantly, in good agreement with what we observed in the weakly textured RVEs ($\theta_{\mathrm{max}}=45^\circ$) presented in Figure \ref{fig:size_low}. However, the grain-size effect on the elastic stiffness and inelastic strength in RVEs with the same grain boundary strength (or GB connectivity with $d_{\mathrm{cr}}=0.5a$) is found to nearly vanish when loaded at $\psi=45^\circ$, as experimentally and numerically evidenced in Figure \ref{fig:size_high}e1. These results reveal that the degree of connectivity throughout the GB regions does not sufficiently account for the grain-size effects in these strongly textured polycrystals. The loading direction ($\psi$)-dependent grain-size effects in the strongly textured polycrystalline RVEs are further addressed. Figures \ref{fig:size_high}d2 and \ref{fig:size_high}e2 present the distributions of $\varphi$, an angle between the crystal orientation and the loading direction ($\varphi \in [0^\circ,45^\circ]$ due to the symmetry) and the directional elastic modulus in each grain (see Figure \ref{fig:single_grain}b in Appendix \ref{section:microstructural} for details). The loading direction significantly influences the directional elastic modulus across the strongly textured grains in these RVEs; i.e., when loaded at $\psi=0^\circ$, as demonstrated in Figure \ref{fig:size_high}d2, most of the grains (or crystal orientations) are found to be not aligned with the loading direction, resulting in low elastic moduli across the grains, whereas they are aligned well with the loading direction when $\psi=45^\circ$. This leads to the relatively high elastic modulus in each of the grains shown in Figure \ref{fig:size_high}e2; i.e., when loaded at $\psi=45^\circ$, the elastic stiffening (or inelastic strengthening) effect in the RVEs with smaller grains ($d=3.75$) due to the higher volume fraction of the GB regions significantly diminishes due to the high stiffness (or strength) in the grain interiors (GI). This is further evidenced in the strongly textured RVEs with low GB connectivity ($d_{\mathrm{cr}}=0.8a$). As shown in the lower inset of Figure \ref{fig:size_high}c, when loaded at $\psi=45^\circ$, the polycrystalline RVEs with larger grains ($d=15$) have much higher elastic moduli compared to those with $d=3.75$; i.e., the “inverse” grain-size effect emerges in these strongly textured architected polycrystals with low GB strength. Polycrystalline metals with extremely fine or nano-sized grains have often been found to exhibit an “inverse” Hall-Petch relationship, e.g., a more softened yield strength with a decreased grain size \cite{argon2006strongest,meyers2006mechanical,carlton2007behind,cordero2016six}. However, a continuous strengthening or hardening with a decreased grain size (i.e., increased GB volume fraction) can emerge in nanocrystalline materials where such GB-mediated deformation modes as grain boundary sliding or rotation are suppressed, as recently demonstrated in recent studies that improved the stability or cohesive strength in the GB regions \cite{rupert2012grain,ozerincc2012grain,khalajhedayati2014emergence,hu2017grain,zhou2020high}. Translating these new physical metallurgical mechanisms, we herein have demonstrated that the interplay between the grain boundary strength and grain interior strength is central to the emergence of grain size-dependent elastic and inelastic features in architected polycrystals. More specifically, our architected materials with greater GB strength are found to exhibit more pronounced grain-size effects, analogous to polycrystalline metals with enhanced resistance against the GB-mediated deformations.

\section{Energy dissipation, local failure and reusability}
\label{Energy dissipation, local failure and reusability}
\begin{figure}[t!]
    \centering
    \includegraphics[width=1.0\textwidth]{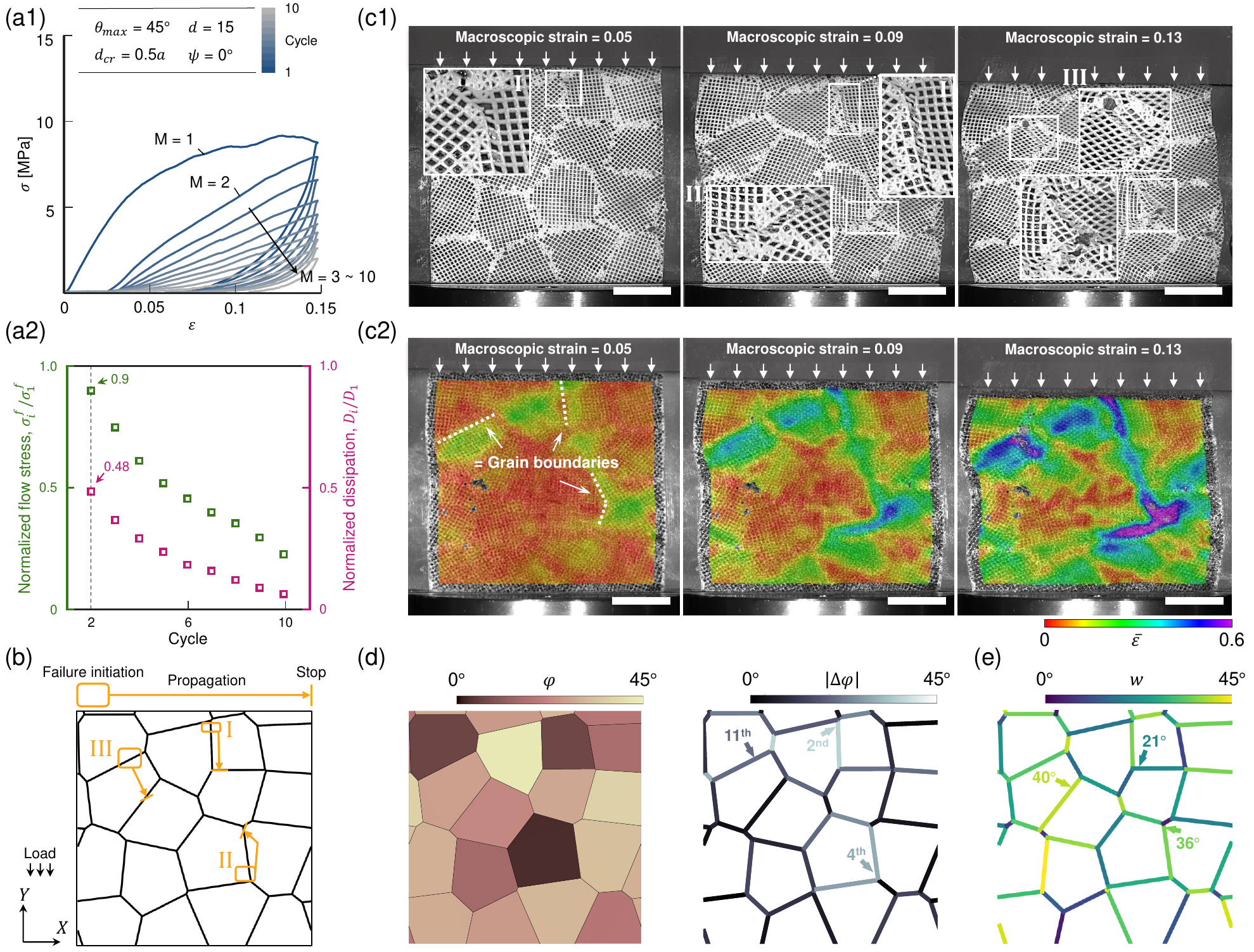}
    \caption{Reusability of weakly textured ($\theta_{\mathrm{max}}=45^\circ$) polycrystals. (a1) Stress-strain curves in polycrystalline RVEs with $d=15$ and $d_{\mathrm{cr}}=0.5a$ loaded at $\psi=0^\circ$ during multiple loading and unloading cycles of $M = 1$ to 10 at a strain rate of 0.01$\mathrm{s}^{-1}$. (a2) Normalized flow stress at the maximum imposed strain of 0.15 and normalized energy dissipation relative to those when $M = 1$. (b) Schematic of representative failure processes, denoted I, II and III, during $M = 1$. Snapshots of (c1) experimental images (insets: magnified images of local failures, I, II and III) and (c2) the corresponding von Mises strain fields (DIC) at increasing macroscopic strains of 0.05, 0.09 and 0.13 during $M = 1$. (d) Distributions of $\varphi$ and $|\Delta \varphi|$ in neighboring grain pairs; the grain boundaries where the local failures (I, II and III) “initiate” are highlighted. (e) Distribution of disorientation angles $\omega$; the grain boundaries where the local failure propagation “stops” are also highlighted. Scale bars: 3cm.}
    \label{fig:reusab_low}
\end{figure}

Next, we examined reusability of the heterogeneous architected polycrystals as a function of “shape recovery” and “energy dissipation” through cyclic loading-unloading mechanical testing; to this end, we followed standard experimental protocols that have been widely used to characterize the resilience and dissipation in elasto-plastomeric materials \cite{greviskes2010effects,cho2013constitutive,cho2017deformation}. Also, the idling time for shape recovery between cycles from $M = 1$ to 10 was 1 hour (see Appendix \ref{section:mechanical}). Figure \ref{fig:reusab_low}a1 presents the stress-strain curves in polycrystalline RVEs with $\theta_{\mathrm{max}}=45^\circ$ loaded at $\psi=0^\circ$; here, the grain size and the critical distance for GB connectivity were set to $d=15$ and $d_{\mathrm{cr}}=0.5a$, correspondingly; i.e., we probe the resilience and dissipation in the RVEs with large grains as well as high GB connectivity in multiple cycles. These RVEs exhibit relatively stiff linear elastic behavior, followed by the inelastic stress-rollover (or yield) with strain hardening during loading. Highly nonlinear unloading behavior is then observed, resulting in a significant amount of mechanical energy dissipation as well as a large amount of “residual” strain; upon unloading, the RVEs recover their original shape by $\sim$ 66\% relative to the residual strain (i.e., from 0.074 to 0.025) at the end of unloading during the idling period (1 hour). Furthermore, the subsequent reloading curves ($M = 2 \sim 10$) show significant degradation not only in the flow stresses at increasing strains but also in the load transfer capability under these harsh conditions of repeated loading and unloading cycles. The experimentally measured stress-strain data are further reduced to the normalized flow stress, especially at the maximum imposed strain of 0.15 and the normalized energy dissipation, relative to those when $M = 1$, as plotted in Figure \ref{fig:reusab_low}a2. Both the flow stress and energy dissipation continue to degrade; in particular, the dissipated energy density is shown to decrease dramatically by $\sim$ 52\% between the first two cycles of $M = 1$ and $M = 2$. The energy dissipation and load transfer capabilities in these RVEs are examined further with no idling period for additional shape recovery between the cycles, as shown in Figure \ref{fig:cyclic} in Appendix \ref{section:reusability}. The significant degradation in the energy dissipation as well as the load transfer capabilities is mainly attributed to failures throughout the hard polycrystalline network, especially during the first cycle. Figure \ref{fig:reusab_low}b schematically illustrates three representative failure processes, denoted I, II and III, observed in this particular RVE ($d=15$ and $d_{\mathrm{cr}}=0.5a$) during $M = 1$. Interestingly, most of the failures are found to initiate at the grain boundary regions, after which they rapidly propagate through the neighboring grains and stop at the other GB regions. The experimental images and the corresponding deformation fields (the von Mises strain fields measured via digital image correlation; see Appendix \ref{section:mechanical}), as plotted in Figures \ref{fig:reusab_low}c1 and \ref{fig:reusab_low}c2, further show these failures in more detail. The first, far-left inset in Figure \ref{fig:reusab_low}c1 magnifies the GB region where the first failure (I) initiates; the middle inset shows the propagation of the first failure (I) through an adjacent grain interior and the initiation of the second failure (II) at a grain boundary region. Lastly, the far-right inset shows the propagation of the second failure (II) through another grain interior; the third failure (III) initiates and propagates throughout another grain boundary (see Movie S2 for details). More importantly, as displayed in Figure \ref{fig:reusab_low}c2, the very large von Mises strains with significant inhomogeneity are localized, especially at the grain boundaries where the three (I, II and III) representative local failures initiate. To address the failure mechanisms at the GB regions in more detail, we plot the distribution of $\varphi$ together with $|\Delta \varphi|$ in neighboring grain pairs in Figure \ref{fig:reusab_low}d. It should be noted that the three failures, denoted I, II and III, are found to initiate at the grain boundaries with a high $|\Delta \varphi|$; more specifically, local failures initiate at the GBs with the second, fourth and eleventh highest values of $|\Delta \varphi|$ among the total 48 grain pairs within the RVE, revealing that the deformation inhomogeneity across the grains gives rise to such local failures. This is further supported by the von Mises stress contours at the GBs with the high $|\Delta \varphi|$ shown in Figure \ref{fig:stress_concentration} in Appendix \ref{section:stress}; a very large stress concentration occurs in the GB region with the second highest $|\Delta \varphi|$, where the first failure, denoted I, initiates at the very early stage of loading during the first cycle, $M = 1$. These results demonstrate that in weakly textured polycrystalline RVEs ($\theta_{\mathrm{max}}=45^\circ$), the initiation of failures is primarily governed by the inter-grain deformation inhomogeneity accommodated at the GB regions. We subsequently explore where the local failure propagation stops throughout the polycrystalline network within these low textured RVEs. Figure \ref{fig:reusab_low}e shows the disorientation angle $\omega$ in each of the GB regions; specifically, disorientation angles ($\omega=21^\circ, 40^\circ$ and $36^\circ$) are displayed at the GBs where the three local failures (denoted I, II and III in Figures \ref{fig:reusab_low}b and \ref{fig:reusab_low}c1) do not propagate further. We note that all of these regions, where I, II and III stopped, are high-angle grain boundaries ($\omega>15^\circ$). In other words, the significant change in the crystal orientation across the grain boundaries can greatly help to mitigate catastrophic failures in these architected polycrystalline materials.

\begin{figure}[t!]
    \centering
    \includegraphics[width=1.0\textwidth]{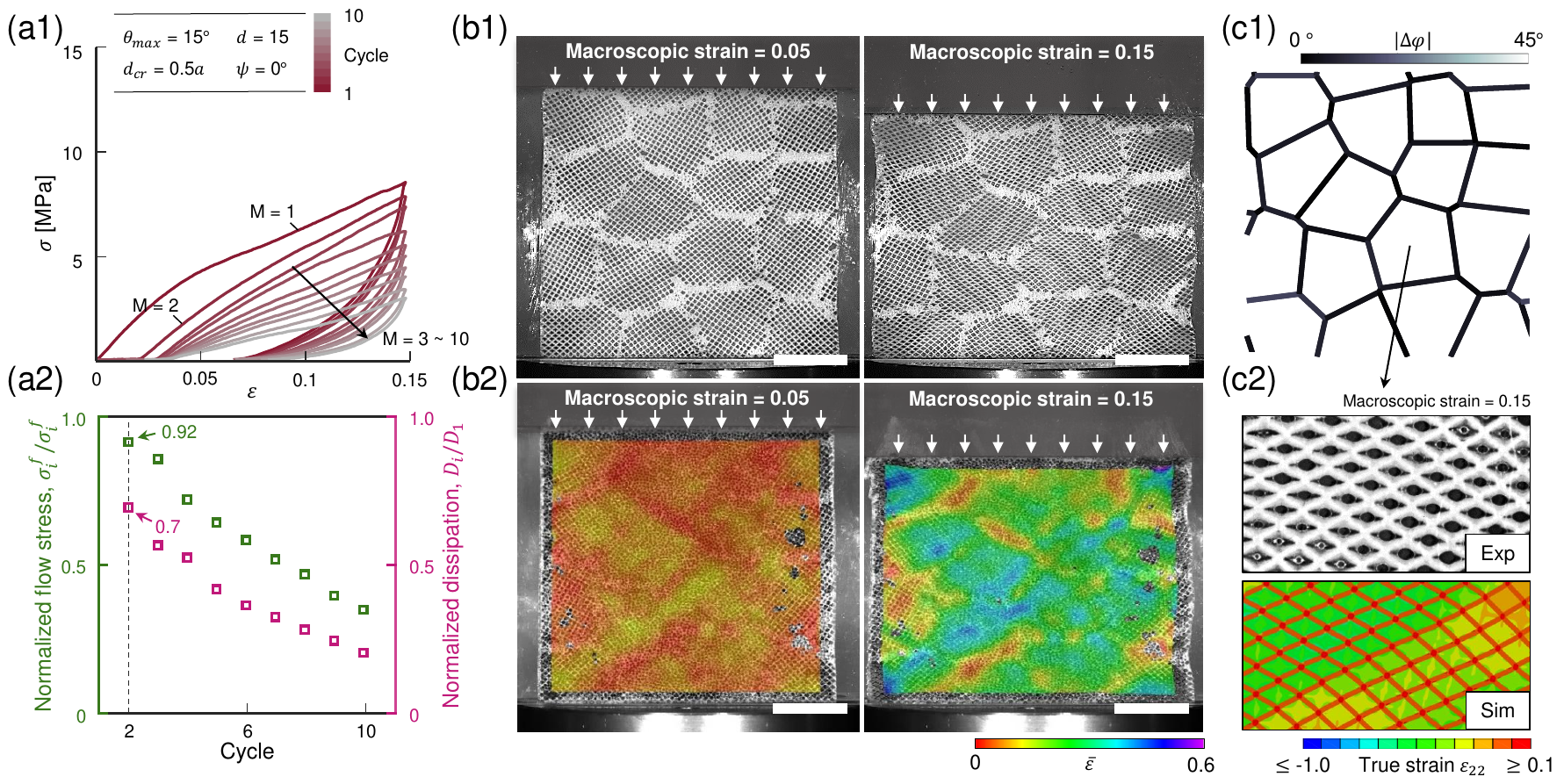}
    \caption{Reusability of strongly textured ($\theta_{\mathrm{max}}=15^\circ$) polycrystals with $d=15$ and $d_{\mathrm{cr}}=0.5a$ loaded at $\psi=0^\circ$. (a1) Stress-strain curves during multiple loading and unloading cycles of $M = 1$ to 10 at a strain rate of 0.01$\mathrm{s}^{-1}$. (a2) Normalized flow stress at the maximum imposed strain of 0.15 and normalized energy dissipation relative to those when $M = 1$. Snapshots of (b1) experimental images and (b2) the corresponding von Mises strain fields (DIC) at increasing macroscopic strains of 0.05 and 0.15 during $M = 1$. (c1) Distribution of $|\Delta \varphi|$ in neighboring grain pairs and (c2) deformed configurations of a representative grain interior. Scale bars: 3cm.}
    \label{fig:reusab_high_0deg}
\end{figure}

\begin{figure}[t!]
    \centering
    \includegraphics[width=1.0\textwidth]{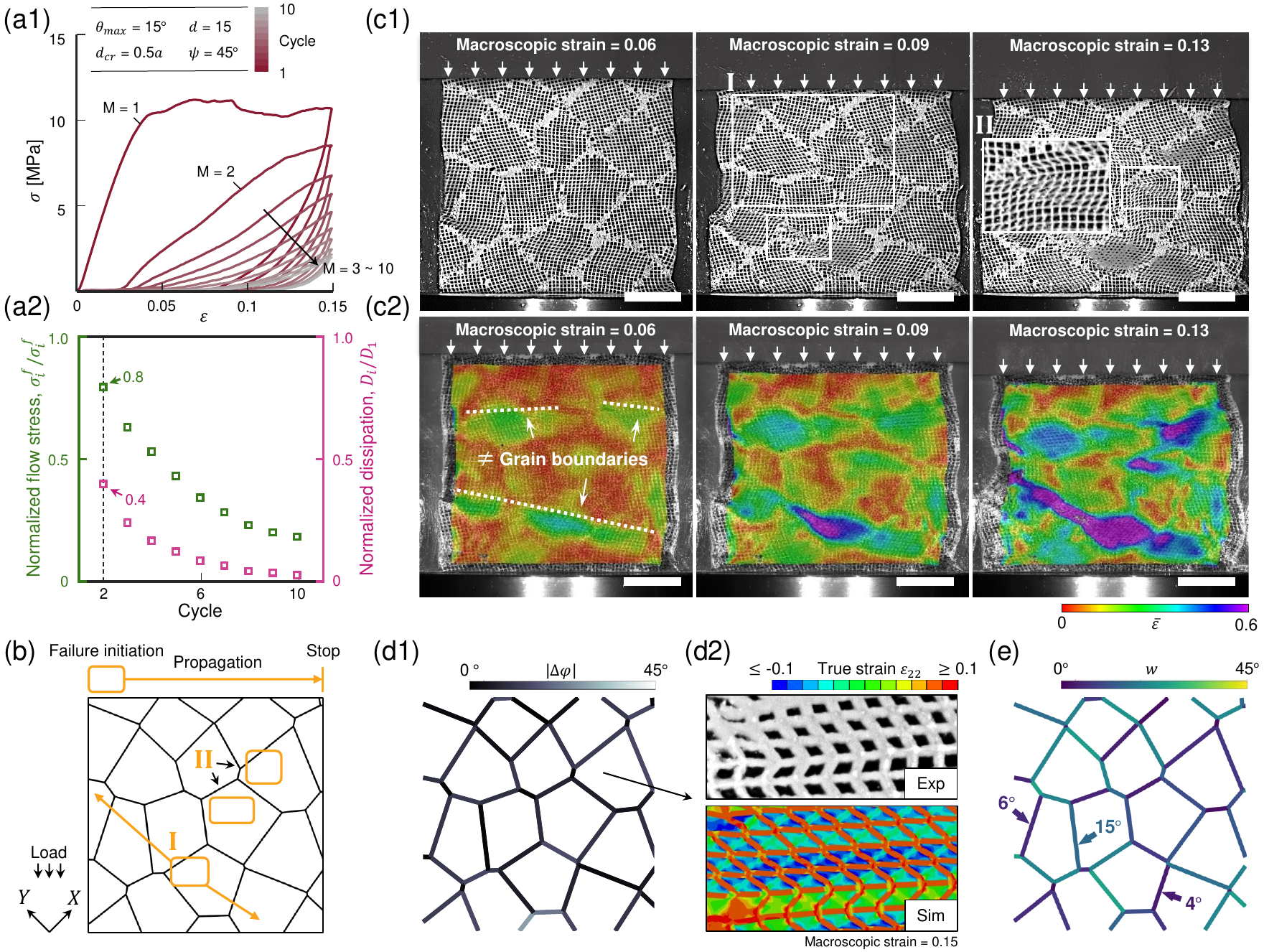}
    \caption{Reusability of strongly textured ($\theta_{\mathrm{max}}=15^\circ$) polycrystals with $d=15$ and $d_{\mathrm{cr}}=0.5a$ loaded at $\psi=45^\circ$. (a1) Stress-strain curves during multiple loading and unloading cycles of $M = 1$ to 10 at a strain rate of 0.01$\mathrm{s}^{-1}$. (a2) Normalized flow stress at the maximum imposed strain of 0.15 and normalized energy dissipation relative to those when $M = 1$. (b) Schematic of representative failure processes, denoted I and II, during $M = 1$. Snapshots of (c1) experimental images (insets: magnified images of local failures, I and II) and (c2) the corresponding von Mises strain fields (DIC) at increasing macroscopic strains of 0.06, 0.09 and 0.13 during $M = 1$. (d1) Distribution of $|\Delta \varphi|$ in neighboring grain pairs and (d2) deformed configurations of a representative grain interior. (e) Distribution of disorientation angles $\omega$; the grain boundaries “unable to stop” the local failure propagation are highlighted. Scale bars: 3cm.}
    \label{fig:reusab_high_45deg}
\end{figure}

Last, we examined the reusability of the heterogeneous architected polycrystals with a strong texture ($\theta_{\mathrm{max}}=15^\circ$), as shown in Figure \ref{fig:reusab_high_0deg} (RVEs loaded at $\psi=0^\circ$) and Figure \ref{fig:reusab_high_45deg} (RVEs loaded at $\psi=45^\circ$). Figures \ref{fig:reusab_high_0deg}a1 and \ref{fig:reusab_high_45deg}a1 present the stress-strain curves of RVEs loaded at $\psi=0^\circ$ and $\psi=45^\circ$, where strongly loading direction ($\psi$)-dependent cyclic behaviors are observed. The RVEs loaded at $\psi=45^\circ$ exhibit much greater elastic stiffness and inelastic strength than those loaded at $\psi=0^\circ$, as the crystal orientations in most of the grains are more aligned with the loading direction in the RVEs loaded at $\psi=45^\circ$; see Figures \ref{fig:size_high}d2 and \ref{fig:size_high}e2 of the distributions of $\varphi$ when loaded at $\psi=0^\circ$ and $45^\circ$. Note that $\varphi$ was defined as the angle between the local crystal orientation and the loading direction. Importantly, upon unloading, energy dissipation and residual strain are found to be greater in the RVEs loaded at $\psi=45^\circ$, followed by significant degradation in the load transfer capability in the subsequent reloading cycles; during $M = 1$, the dissipated energy density and residual strain are $570 \mathrm{kJm}^{-3}$ and 0.068 in a RVE loaded at $\psi=0^\circ$, and are $1189 \mathrm{kJm}^{-3}$ and 0.077 in a RVE loaded at $\psi=45^\circ$. Furthermore, the RVEs loaded at $\psi=0^\circ$ exhibit moderate decreases in the flow stress and energy dissipation of $\sim$ 8\% and $\sim$ 30\%, respectively, between $M = 1$ and $M = 2$, with gradual degradation observed from $M = 2$ to $M = 10$, as displayed in Figure \ref{fig:reusab_high_0deg}a2. Especially when $M = 10$, the RVEs loaded at $\psi=0^\circ$ still show relatively decent load transfer and energy dissipation capabilities of correspondingly $\sim$ 35\% and $\sim$ 20\% relative to those when $M = 1$. However, as depicted in Figure \ref{fig:reusab_high_45deg}a2, the flow stress and energy dissipation in the RVEs loaded at $\psi=45^\circ$ are found to decrease substantially by $\sim$ 20\% and $\sim$ 60\% between $M = 1$ and $M = 2$ and stabilize at $M \geq 7$. Finally, when $M = 10$, the load transfer and energy dissipation capabilities are shown correspondingly to degrade by $\sim$ 72\% and $\sim$ 98\%, relative to those at $M = 1$. The RVEs loaded at $\psi=0^\circ$ exhibit the relatively high reusability since there are no significant local failures throughout the polycrystalline network during the first cycle, $M = 1$, as displayed in Figure \ref{fig:reusab_high_0deg}b1. This is further evidenced by the highly uniform strain distribution with no significant localization, as presented in Figure \ref{fig:reusab_high_0deg}b2 (see Movie S3 for details). In contrast, the RVEs loaded at $\psi=45^\circ$ are found to undergo multiple local failures. The first failure (I) initiated within a grain “interior” and propagated throughout the neighboring grain network, followed by subsequent local failures (II), as schematically shown in Figure \ref{fig:reusab_high_45deg}b. These local failure processes (I and II) are also highlighted in the experimental images together with the corresponding DIC deformation fields in Figures \ref{fig:reusab_high_45deg}c1 and \ref{fig:reusab_high_45deg}c2 (see Movie S4 for details); the middle inset of Figure \ref{fig:reusab_high_45deg}c1 shows the initiation and propagation of the first failure (I) at the macroscopic strain of 0.09, and the far-right inset clearly shows the two local failures (II) simultaneously emanating from the grain interiors at the macroscopic strain of 0.13, as well evidenced in the corresponding DIC images plotted in Figure \ref{fig:reusab_high_45deg}c2. We note that these local failures initiate within the “grain interiors”, not at the grain boundaries. It should also be noted that $|\Delta \varphi|$ is relatively small throughout the grain boundary networks in these RVEs with a strong texture ($\theta_{\mathrm{max}}=15^\circ$) loaded at both $\psi=0^\circ$ (Figure \ref{fig:reusab_high_0deg}c1) and $\psi=45^\circ$ (Figure \ref{fig:reusab_high_45deg}d1); the highest $|\Delta \varphi|$ is found to be $\sim 14^\circ$ in these RVEs. In other words, the neighboring grain pairs throughout these RVEs with a strong texture are less likely to experience significant deformation “inhomogeneity” across the grain boundary regions, as is well evidenced in the far-left inset in Figure \ref{fig:reusab_high_45deg}c2. These results imply that the GB regions do not play a crucial role in the initiation of local failures in these RVEs with a strong texture while their role is critical in weakly textured architected polycrystals with $\theta_{\mathrm{max}}=45^\circ$, as demonstrated in Figure \ref{fig:reusab_low}.

The microstructures in each of the grains further reveal the underlying failure mechanisms in the strongly textured polycrystalline RVEs. When loaded at $\psi=0^\circ$, the crystal orientations in the grain interiors are not aligned with the loading direction, as evidenced by the distribution of $\varphi$ (the angle between the crystal orientation and the loading direction) in Figure \ref{fig:size_high}d2, by which the textured RVEs can effectively accommodate the imposed deformation predominantly through “bending” in the hard struts with no significant local failures. This is well evidenced in Figure \ref{fig:reusab_high_0deg}c2, which displays deformed configurations of a representative grain interior from both an experiment (upper inset) and a numerical simulation (lower inset). In particular, the lower inset of Figure \ref{fig:reusab_high_0deg}c2 shows a relatively homogeneous strain field throughout the soft domains via bending-dominated deformation in the hard strut network. However, as shown in Figure \ref{fig:reusab_high_45deg}d2 of a “deformed” grain interior in the RVE loaded at $\psi=45^\circ$, the grain interior microstructures are found to buckle; the buckling culminates in a catastrophic failure in the grain interior, as is well evidenced in the numerical simulation result that illustrates the pronounced localization of deformation in the lower inset of Figure \ref{fig:reusab_high_45deg}d2. Furthermore, as clearly displayed in the series of DIC images in the insets of Figure \ref{fig:reusab_high_45deg}c2, the first failure, denoted I, is shown to propagate rapidly throughout the neighboring grains (see the far-right inset; a shear band with an extremely high Mises strain $>$ 0.6 is observed across at least six grains). It should also be noted that the grain boundaries between these “failed” grains are found to be unable to stop the rapid, catastrophic failure propagation. As shown in Figure \ref{fig:reusab_high_45deg}e, the disorientation angles $\omega$ (the minimum angle required to perfectly align two adjacent crystal lattices) are found to be not greater than $15^\circ$. This suggests that the low-angle GBs ($\omega<15^\circ$) hardly impede failure propagation, in good agreement with the results of the weakly textured RVEs presented in Figure \ref{fig:reusab_low}e. Once the local failure due to buckling or shear band initiates, it leads not only to the inevitable catastrophic failure throughout the entire RVE but also to significant degradation of the load transfer and energy dissipation capabilities during the subsequent loading and unloading cycles.

Our heterogeneous architected polycrystals with strong texture ($\theta_{\mathrm{max}}=15^\circ$, Figures \ref{fig:reusab_high_0deg} and \ref{fig:reusab_high_45deg}) exhibited the severe strain localization within grain interiors especially when loaded at $\psi=45^\circ$ since the struts were aligned well with the loading direction. Similarly, such strain localizations or shear bands are often observed in strongly textured polycrystalline metals when the crystal orientation in each of the grains is favorable for slips \cite{echlin2016incipient,hemery20193d,cappola2021localization}. Additionally, strain localizations are known to easily propagate across the boundaries between grains with well-aligned crystal orientations; i.e., a low disorientation angle, eventually leading to the emergence of long-range slip transmissions \cite{echlin2016incipient,hemery20193d,stinville2023insights} or macroscopic shear bands \cite{anand1994process} in polycrystalline metals. By contrast, a significant change in the crystal orientations across grain boundaries has been found to increase resistance against slip or shear band transmissions \cite{sangid2013physics,imrich2014differences,dehm2022implication}; furthermore, polycrystalline metals with randomly distributed crystal orientations are widely known to exhibit dominant intra-grain slip activities and dislocation pile-ups at grain boundaries. Importantly, as demonstrated in the weakly textured architected polycrystals ($\theta_{\mathrm{max}}=45^\circ$) in Figure \ref{fig:reusab_low}, the high-angle GBs ($\omega>15^\circ$) are found to effectively hinder the propagation of local failures into neighboring grains. These similarities suggest that visualizing the inter- and intra-grain deformation processes in our “macroscopic” architected polycrystals helps to better understand the extreme deformations in polycrystalline metals despite the differences in underlying deformation mechanisms and length scales between the architected materials and their atomic counterparts.

\section{Conclusion}
\label{Conclusion}
We have presented and demonstrated the heterogeneous architected materials by a combination of micromechanical modeling, experiments and numerical simulations, inspired by metallic polycrystals comprising randomly oriented grains and their coherent boundaries. The heterogeneous architected polycrystals have been shown to exhibit outstanding combinations of stiffness, yield strength, flow stresses, energy dissipation and shape recovery strongly dependent upon both the grain sizes and textures, analogous to those in polycrystalline metals. The interplay between the grain boundaries and grain interiors gives rise to the grain-size effects on elastic and inelastic features in these architected polycrystals. Importantly, through the careful design of connectivity and strength throughout the grain boundary network, the grain-size effects have been demonstrated in our 3D-printed polycrystal prototypes. Furthermore, by visualizing the initiation and propagation of inter- and intra-grain failures throughout the polycrystalline networks, we also have shown that the crystallographic texture plays a crucial role in reusability of the architected polycrystals subjected to severe deformation conditions.

The striking analogy between our architected polycrystals and their atomic counterparts demonstrated in this work reveals the universality of grain boundary and texture engineering in this new class of heterogeneous materials. Furthermore, it allows us to better understand the complex deformation mechanisms in a broad class of polycrystalline materials with randomly oriented grains and the boundaries. Finally, not limited to those presented in this work, our approach will lead to a new emergence of heterogeneous architected polycrystals with more realistic grain boundary geometry and topology in a three- or four-dimensional setting across a wide range of length scales.

\section*{Acknowledgement}
We acknowledge financial support from the National Research Foundation of Korea through Grant No. RS-2023-00279843.
\clearpage

\setcounter{figure}{0}
\setcounter{table}{0}
\setcounter{equation}{0}
\setcounter{section}{0}

\renewcommand{\theHfigure}{A.\arabic{figure}}
\renewcommand{\theHtable}{A.\arabic{table}}
\renewcommand{\theHequation}{A.\arabic{equation}}

\appendix
\renewcommand{\thesection}{Appendix~\Alph{section}}
\labelformat{section}{\Alph{section}}

\section{Microstructural details in heterogeneous architected polycrystals}
\label{Microstructural details in heterogeneous architected polycrystals}
\renewcommand{\thefigure}{A.\arabic{figure}}
\renewcommand{\thetable}{A.\arabic{table}}
\renewcommand{\theequation}{A.\arabic{equation}}
\renewcommand{\thesubsection}{A.\arabic{subsection}}
\subsection{Constructing grain boundary regions}
\label{section:constructing}
\begin{figure}[b!]
    \centering
    \includegraphics[width=0.8\textwidth]{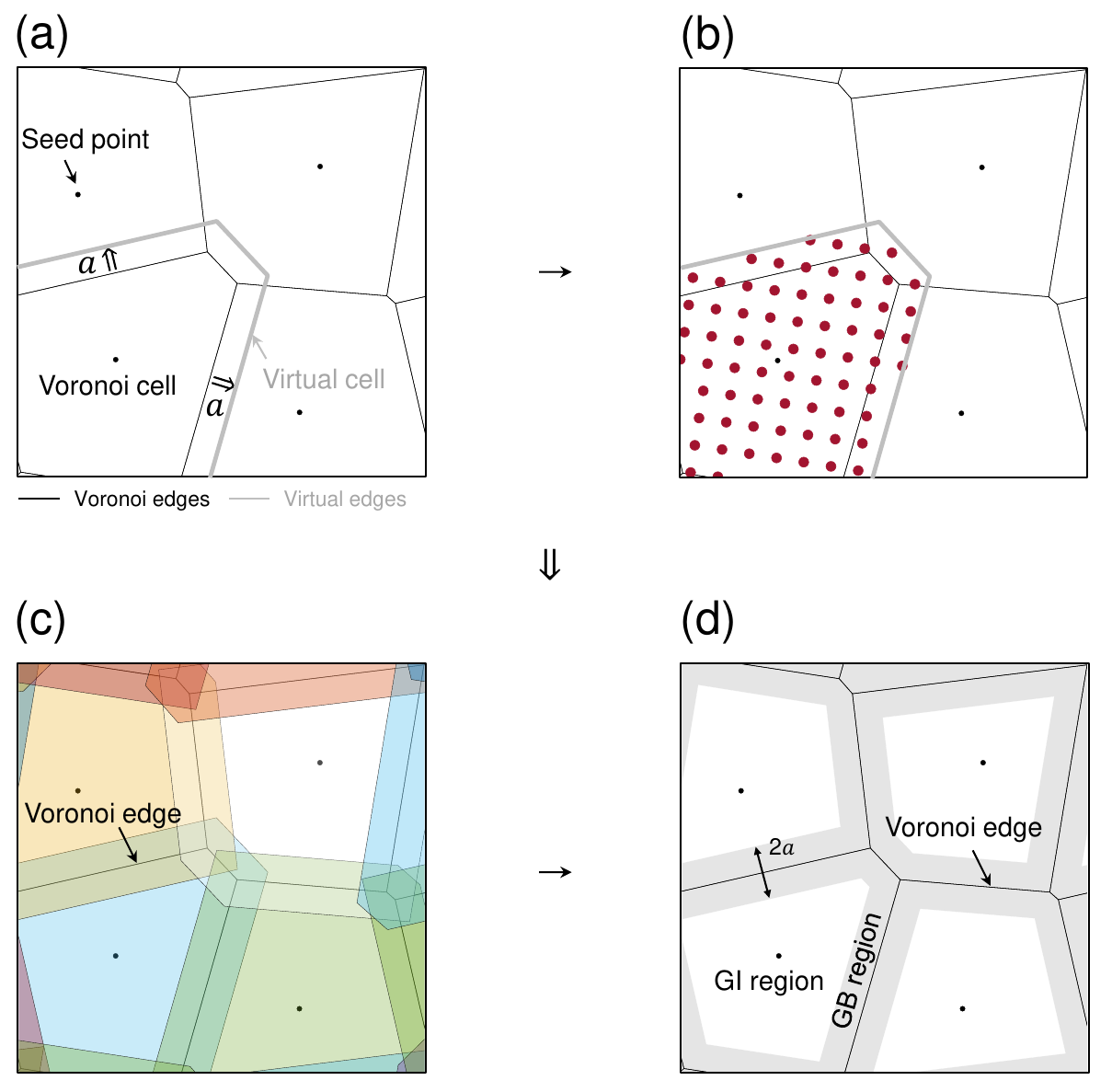}
    \caption{A procedure for constructing grain boundary regions in heterogeneous architected polycrystals. (a) A virtual cell created by expanding a Voronoi cell by a lattice constant $a$, (b) the maximally arranged atoms within the virtual cell, (c) a schematic representation of virtual cells; each virtual cell is displayed in a different color, and (d) resulting polycrystalline domains divided into grain interiors and grain boundary regions with a thickness of 2$a$.}
    \label{fig:gb_design}
\end{figure}
Grain boundary (GB) regions in heterogeneous architected polycrystals proposed in this work are constructed, as follows. As shown in Figure \ref{fig:gb_design}a, each of the Voronoi cells generated from $N$ seed points, randomly distributed in an $L \times L$ domain, is expanded by a lattice constant $a$, thus resulting in a slightly larger polygon defined as a virtual cell. Note that individual grains are constrained to grow within the virtual cells; for example, Figure \ref{fig:gb_design}b displays the maximally arranged atoms (in red) within a virtual cell during the simple crystal growth. As a result, the $L \times L$ domain consists of the virtual cells, each enclosed by the virtual edges, as schematically shown in Figure \ref{fig:gb_design}c. Then, all overlapping regions of the virtual cells (i.e., the regions enclosed by the virtual edges) are taken to be the GB regions; the remainder are grain interior (GI) regions, as shown in Figure \ref{fig:gb_design}d.

\subsection{Tailoring connectivity (or strength) throughout grain boundary networks}
\label{section:tailoring}
\begin{figure}[b!]
    \centering
    \includegraphics[width=1.0\textwidth]{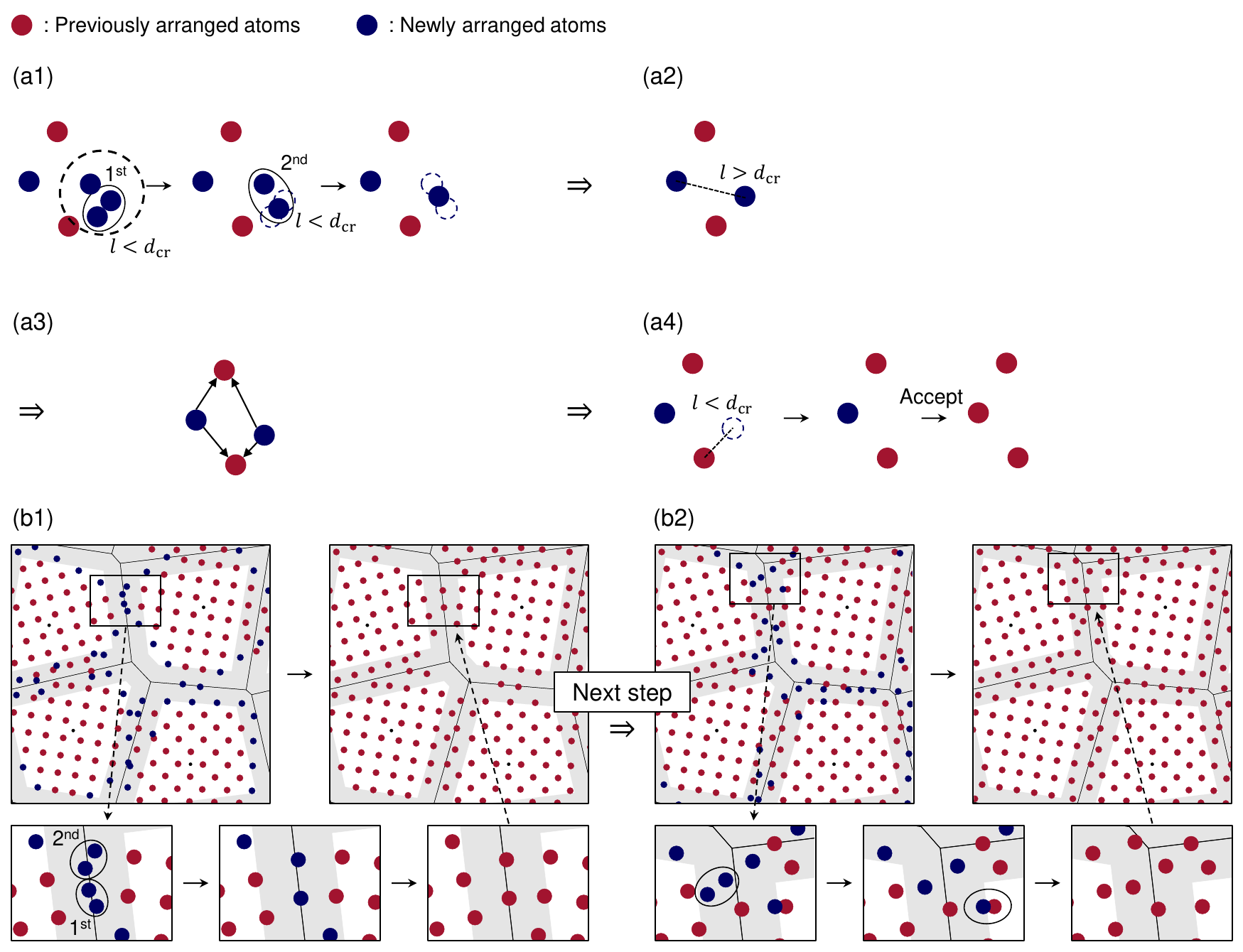}
    \caption{Arranging atoms in the grain boundary regions. (a) Rules of merging and removing newly arranged atoms. (b1) and (b2) present examples of merging and removing atoms in a grain boundary region during the construction of an heterogeneous architected polycrystal demonstrated in this work.}
    \label{fig:gb_strength}
\end{figure}
We here present a detailed procedure for merging and removing the neighboring atomic pairs whose distances are shorter than $d_{\mathrm{cr}}$ (i.e., $l < d_{\mathrm{cr}}$) in the grain boundary regions. At each step of the radial arrangement process, atoms may be identified to be either previously arranged (in red) or newly arranged (in blue). For example, let us consider a set of previously arranged (in red) and newly arranged (in blue) atoms, as shown in Figure \ref{fig:gb_strength}a1. If the distance between a pair of neighboring, newly arranged atoms (in blue) is shorter than $d_{\mathrm{cr}}$, the atoms are simply merged into a new atom (in blue) located at the midpoint, as illustrated in the lower inset of Figure \ref{fig:design}b in the main text. However, in most cases, the atoms are arranged such that the distances between multiple pairs of neighboring, newly arranged atoms (in blue) are shorter than $d_{\mathrm{cr}}$, as schematically displayed in Figure \ref{fig:gb_strength}a1; i.e., the distances between all possible pairs of the multiple, newly arranged atoms (within the dashed circle) may be all shorter than $d_{\mathrm{cr}}$. In this case, the closest atomic pair is merged first, and the next closest pairs are sequentially merged into a new atom at the midpoint. This process continues until no more atoms remain to be merged (i.e., $l < d_{\mathrm{cr}}$), as shown in Figure \ref{fig:gb_strength}a2. Then, as illustrated in Figure \ref{fig:gb_strength}a3, the distances between the previously arranged atoms (in red) and the newly arranged atoms (in blue) are calculated. All of the newly arranged atoms (in blue) within the critical distance from the previously arranged atoms (in red) are removed and only the remaining atoms are accepted at each step as shown in Figure \ref{fig:gb_strength}a4. An example of arranging atoms in the grain boundary regions is illustrated in Figures \ref{fig:gb_strength}b1 and \ref{fig:gb_strength}b2. Since newly arranged atoms (in blue) are less removed or merged as the critical distance $d_{\mathrm{cr}}$ decreases, more atoms are accepted at each step and embedded in the grain boundary regions, by which the grain boundary networks can have greater connectivity (or strength); i.e., the more atoms embedded, the more strengthened the grain boundary regions are.
\clearpage

\subsection{Crystal orientation angle $\theta$, disorientation angle $\omega$, and loading direction $\psi$}
\label{section:microstructural}
\begin{figure}[!h]
    \centering
    \includegraphics[width=1.0\textwidth]{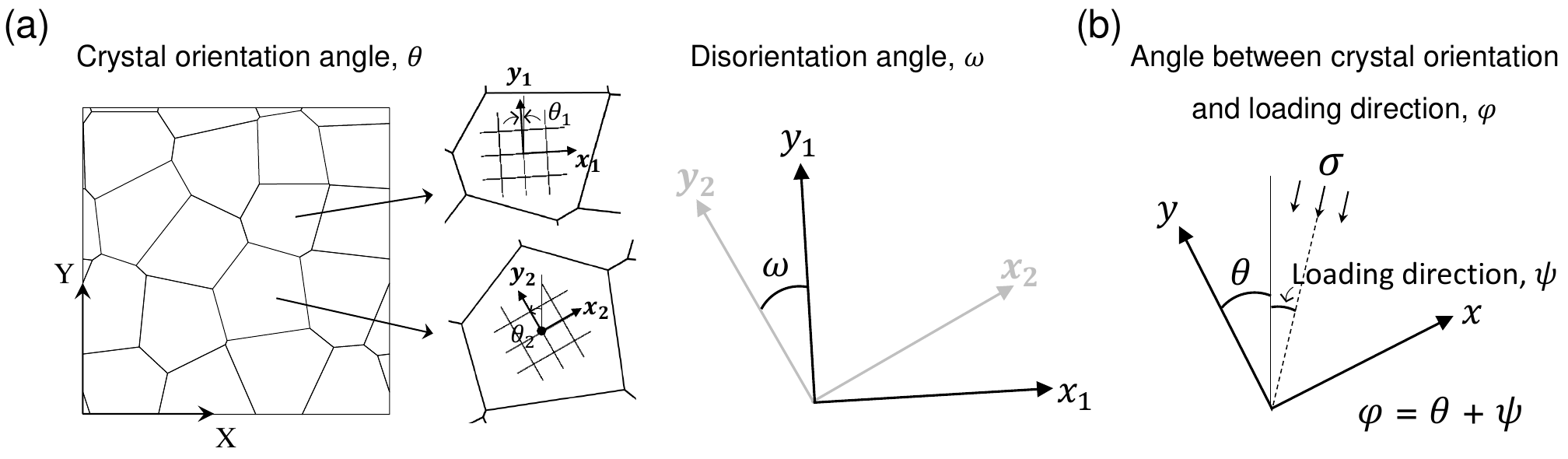}
    \caption{(a) Crystal orientation angle, $\theta$ in each grain and disorientation angle, $\omega$ between two adjacent crystal lattices (or grains). (b) $\varphi$ defined as an angle between the crystal orientation ($\theta$) and the loading direction ($\psi$).}
    \label{fig:single_grain}
\end{figure}

A crystal orientation angle $\theta$ in each grain is randomly selected from a range of $[45^\circ - \theta_{\mathrm{max}},45^\circ + \theta_{\mathrm{max}}]$, as shown in the far left inset of Figure \ref{fig:single_grain}a. A disorientation angle $\omega$ is then calculated as the minimum angle required to align two adjacent or neighboring crystal lattices perfectly, as shown in the far right inset of Figure \ref{fig:single_grain}a. Furthermore, as illustrated in Figure \ref{fig:single_grain}b, by introducing a loading direction $\psi$, we also define $\varphi$ as an angle between the local crystal orientation and the loading direction; note that the loading direction $\psi$ is defined clockwise from the global $Y$-axis of a sample coordinate system $(X,Y)$. Since the local crystal coordinate system $(x,y)$ is rotated counterclockwise by the crystal orientation angle $\theta$ from the $Y$-axis of the global sample coordinate system $(X,Y)$, $\varphi$ is simply given by $\varphi=\theta+\psi$.
\clearpage

\section{Experimental and numerical procedures}
\label{Experimental and numerical procedures}
\renewcommand{\thefigure}{B.\arabic{figure}}
\renewcommand{\thetable}{B.\arabic{table}}
\renewcommand{\theequation}{B.\arabic{equation}}
\renewcommand{\thesubsection}{B.\arabic{subsection}}
\subsection{3D-printed polycrystal prototype and experimental setup for plane strain compression}
\label{section:prototype}
\begin{figure}[!h]
    \centering
    \includegraphics[width=0.95\textwidth]{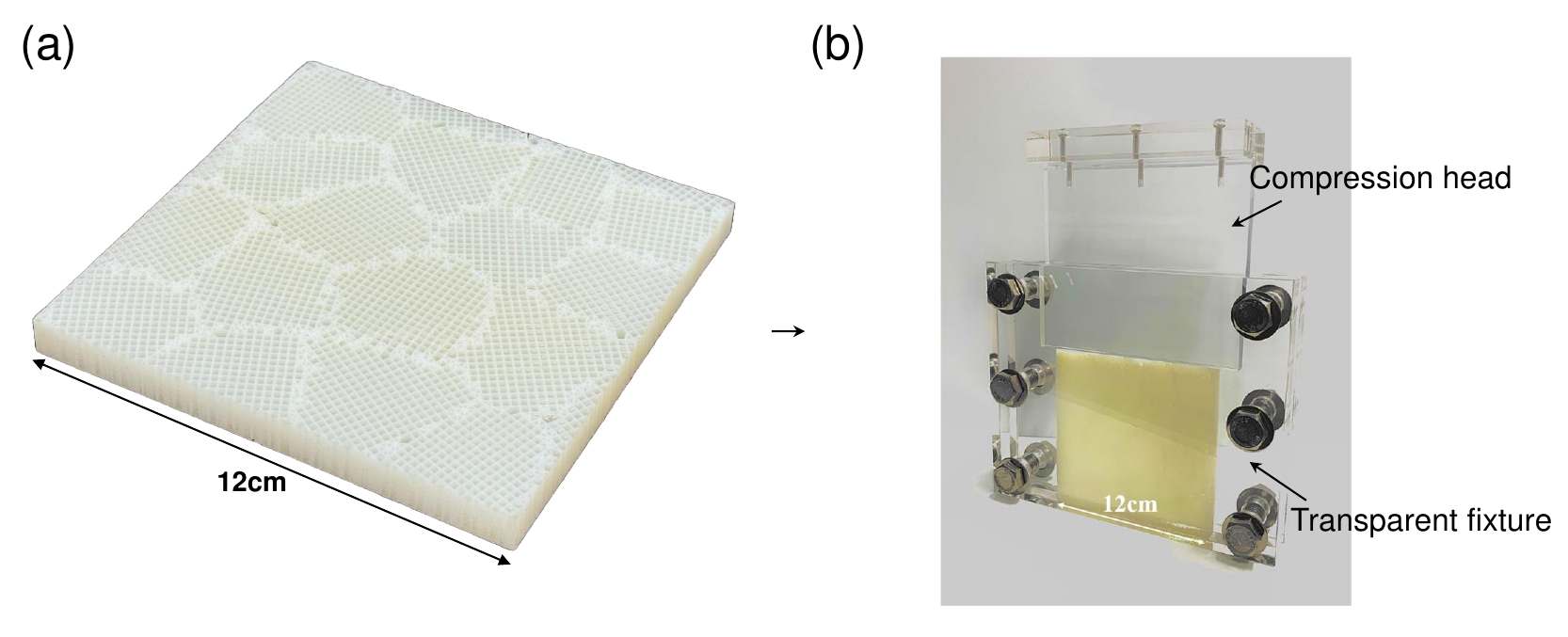}
    \caption{(a) A 3D-printed sample of a weakly textured ($\theta_{\mathrm{max}}=45^\circ$) polycrystalline RVE with $d=15$ and $d_{\mathrm{cr}}=0.5a$ (used in the results presented in Figure \ref{fig:size_low}). A “hard” polycrystalline network (bright region in the sample) is fabricated using a thermoplastic polymer (VeroWhitePlus$^{\mathrm{TM}}$), while isolated “soft” domains (dark region) is made of a rubbery polymer (TangoPlus$^{\mathrm{TM}}$). (b) Experimental setup for plane strain conditions.}
    \label{fig:prototype}
\end{figure}

\subsection{Mechanical testing and deformation field analysis}
\label{section:mechanical}
The 3D-printed samples were subjected to a plane-strain condition, for which they were placed between two transparent acrylic fixtures, as shown in Figure \ref{fig:prototype}b. The samples were then compressed between two plates using a universal mechanical testing machine (INSTRON 4482, load-cell: 100kN) at room temperature (295K). Mineral oil was used as a lubricant between the samples and the fixtures for a frictionless condition. Furthermore, multiple cyclic loading-unloading mechanical tests were performed under two loading scenarios: i) with (see Figures \ref{fig:reusab_low}, \ref{fig:reusab_high_0deg} and \ref{fig:reusab_high_45deg}), and ii) without (see Figure \ref{fig:cyclic}) an idling time of 1 hour for inelastic shape recovery between cycles. All the mechanical tests were performed quasi-statically with an engineering strain rate of 0.01 $\mathrm{s}^{-1}$. For Digital Image Correlation (DIC) analysis on the deformation fields, random speckles were patterned on the front surfaces of samples. The speckles were tracked at a rate of 5fps (VicSnap, Correlated Solutions) during the mechanical testing; then the deformation fields were analyzed and obtained (VIC-2D 6, Correlated Solutions).

\subsection{Micromechanical modeling}
\label{section:micromechanical}
Boundary value problems of the polycrystalline RVEs were solved using nonlinear finite element analysis with Abaqus/Standard. To this end, we developed and numerically implemented a finite deformation constitutive model for the hard component (i.e., a thermoplastic polymer), as described in Appendix \ref{section:large}. The soft component (i.e., a rubbery polymer) was modeled as an Arruda-Boyce hyperelastic representation. We used four-node bilinear elements with reduced integration and hourglass control (CPE4R). Furthermore, at least four elements were put through the thickness direction of the “hard” struts in the polycrystalline RVEs.

Periodic boundary conditions were imposed on each of the polycrystalline RVEs to model the macroscopic behaviors. To this end, we employed the fictitious node virtual work method proposed in \cite{danielsson2002three} and \cite{danielsson2007micromechanics}, by which the macroscopic, averaged responses in the RVEs are computed. First, the periodic boundary condition for a pair of material points ($\mathit{A}$ and $\mathit{B}$ on the periodic pair surfaces) is applied to the RVE as follows
\begin{equation}
\mathbf{u}_{A} - \mathbf{u}_{B} = \left( \bar{\mathbf{F}} - \mathbf{I}\right)\left(\mathbf{X}_{A} - \mathbf{X}_{B} \right),
\label{eq:pbc}
\end{equation}
where $\mathbf{u}$ is the local displacement, $\bar{\mathbf{F}}$ is the prescribed macroscopic deformation gradient for the periodic RVE, $\mathbf{I}$ is the second-order identity tensor and $\mathbf{X}$ is the coordinate vector at the undeformed reference configuration. Eq. \ref{eq:pbc} provides kinematic constraints for pairs of the material points on the periodic RVE surfaces. The macroscopic, averaged stress response in the RVE is then computed using the principle of virtual work such that an internal virtual work due to variations is balanced with an external virtual work, as follows
\begin{equation}
\delta W_{\mathrm{int}} = \delta W_{\mathrm{ext}}.
\label{eq:work_balance}
\end{equation}
The internal virtual work can be expressed in terms of the first Piola (engineering) stress and the deformation gradient (or displacement gradient), as follows
\begin{equation}
\delta W_{\mathrm{int}} = \int_{V_0} \mathbf{T}_\mathrm{R}(\mathbf{X}):\delta\mathbf{F}(\mathbf{X}) d V = V_0 \bar{\mathbf{T}}_{\mathrm{R}}:\delta\bar{\mathbf{F}} = V_0 \bar{\mathbf{T}}_{\mathrm{R}}:\delta\bar{\mathbf{H}},
\label{eq:internalwork}
\end{equation}
where $V_0$ is the volume at the reference configuration, $\mathbf{T}_{\mathrm{R}}$ is the local first Piola stress, $\mathbf{F}\equiv \frac{\partial\bm\upvarphi}{\partial\mathbf{X}}$ is the local deformation gradient that maps the material vector ($\mathbf{X}$) at the reference configuration to the spatial vector ($\mathbf{x}=\bm\upvarphi(\mathbf{X},t)$, $\bm\upvarphi$: motion) at the deformed configuration, $\bar{\mathbf{T}}_{\mathrm{R}}$ is the macroscopic first Piola stress and $\bar{\mathbf{H}}=\bar{\mathbf{F}}-\mathbf{I}$ is the macroscopic displacement gradient. The operation (:) denotes the inner product of two tensors. The external virtual work can be written as
\begin{equation}
\delta W_{\mathrm{ext}} = \int_{S_0} \mathbf{T}_\mathrm{R}(\mathbf{X})\mathbf{n}_0 \cdot \delta \mathbf{u}(\mathbf{X}) d S = \int_{S_0} \mathbf{t}_0(\mathbf{X}) \cdot \mathbf{u}(\mathbf{X}) d S,
\label{eq:externalwork}
\end{equation}
where $\mathbf{t}_0 \equiv \mathbf{T}_\mathrm{R}\mathbf{n}_0$ is the surface traction, $\mathbf{n}_0$ is the outward unit vector normal to the surface and $\delta\mathbf{u}$ is the virtual displacement in the reference configuration. For the periodic RVE, the external virtual work can be computed by
\begin{equation}
\delta W_{\mathrm{ext}} = \mathbf{\Phi}:\delta \mathbf{\bm{\eta}},
\label{eq:externalwork2}
\end{equation}
where the components of $\bm{\eta}$ are the generalized degrees of freedom or displacement components of the fictitious nodes and the components of $\mathbf{\Phi}$ are the work conjugate generalized forces corresponding to the components of $\bm{\eta}$. Here, the admissible variation of the generalized degrees of freedom is identical to the variation of the macroscopic displacement gradient; i.e., $\delta \bm{\eta} = \delta \bar{\mathbf{H}}$. By using Eq. \ref{eq:internalwork}, \ref{eq:externalwork} and \ref{eq:externalwork2}, the macroscopic, averaged first Piola stress in the periodic RVE can be computed by
\begin{equation}
\bar{\mathbf{T}}_{\mathrm{R}} = \frac{1}{V_0} \int_{V_0} \mathbf{T}_{\mathrm{R}} (\mathbf{X}) d V = \frac{1}{V_0} \mathbf{\Phi}
\label{eq:avg_piola}
\end{equation}
\clearpage

\section{Mechanical behavior of constituent hard and soft components: experiments and constitutive modeling}
\label{constitutive modeling}
\renewcommand{\thefigure}{C.\arabic{figure}}
\renewcommand{\thetable}{C.\arabic{table}}
\renewcommand{\theequation}{C.\arabic{equation}}
\renewcommand{\thesubsection}{C.\arabic{subsection}}
\label{section:large}
We present finite deformation constitutive modeling framework for the hard and soft components. The constitutive model of the hard component comprises (1) a time-dependent elastic-viscoplastic mechanism (denoted I) and (2) a time-independent hyperelastic mechanism (denoted N). We then define the following basic kinematic fields for the hard component:
\vspace*{0.3in}

\begin{tabular}{l l}
    $\mathbf{F} \defeq \frac{\partial \bm{\upvarphi}}{\partial \mathbf{X}} = \mathbf{F}_\mathrm{I} = \mathbf{F}_\mathrm{N}$ & deformation gradient that maps material points in a reference \\ & to points ($\mathbf{x} = \bm{\upvarphi}(\mathbf{X},t)$; $\bm{\upvarphi}$: motion) in a deformed configuration; \\
    $J \defeq \mathrm{det}(\mathbf{F}) > 0$ & volume change; \\
    $\mathbf{F}_\mathrm{I} = \mathbf{F}^{e}_\mathrm{I} \mathbf{F}^{p}_\mathrm{I}$ & decomposition of $\mathbf{F}_\mathrm{I}$ into elastic ($\mathbf{F}^{e}_\mathrm{I}$) and plastic ($\mathbf{F}^{p}_\mathrm{I}$) parts; \\
    $\mathbf{F}^{e}_\mathrm{I} = \mathbf{R}^{e}_\mathrm{I} \mathbf{U}^{e}_\mathrm{I}$ & polar decomposition of $\mathbf{F}^{e}_\mathrm{I}$ into rotation ($\mathbf{R}^{e}_\mathrm{I}$) and stretch ($\mathbf{U}^{e}_\mathrm{I}$); \\
    $\bar{\mathbf{F}}_\mathrm{N} = J^{-1/3}\mathbf{F}_\mathrm{N}$ & isochoric part of $\mathbf{F}_\mathrm{N}$; \\
    $\bar{\mathbf{B}}_\mathrm{N} = \bar{\mathbf{F}}_\mathrm{N}\bar{\mathbf{F}}_\mathrm{N}^{\top}$ & isochoric left Cauchy-Green tensor. \\
\end{tabular}
\vspace*{0.3in}

\noindent The deformation rate is described by the velocity gradient $\mathbf{L} \defeq \mathrm{grad}\textbf{v}$ which is decomposed into elastic ($\mathbf{L}^{e}_{\mathrm{I}}$) and plastic ($\mathbf{L}^{p}_{\mathrm{I}}$) parts,
\begin{equation}
\begin{aligned}
\mathbf{L} & =\dot{\mathbf{F}}\mathbf{F}^{-1} \\
& =\dot{\mathbf{F}}^{e}_{\mathrm{I}}\mathbf{F}^{e-1}_{\mathrm{I}}+\mathbf{F}^{e}_{\mathrm{I}}\dot{\mathbf{F}}^{p}_{\mathrm{I}}\mathbf{F}^{p-1}_{\mathrm{I}}\mathbf{F}^{e-1}_{\mathrm{I}} \\
& = \mathbf{L}^{e}_{\mathrm{I}}+\mathbf{F}^{e}_{\mathrm{I}}\mathbf{L}^{p}_{\mathrm{I}}\mathbf{F}^{e-1}_{\mathrm{I}}.
\end{aligned}
\end{equation}

\begin{figure}[t!]
    \centering
    \includegraphics[width=1.0\textwidth]{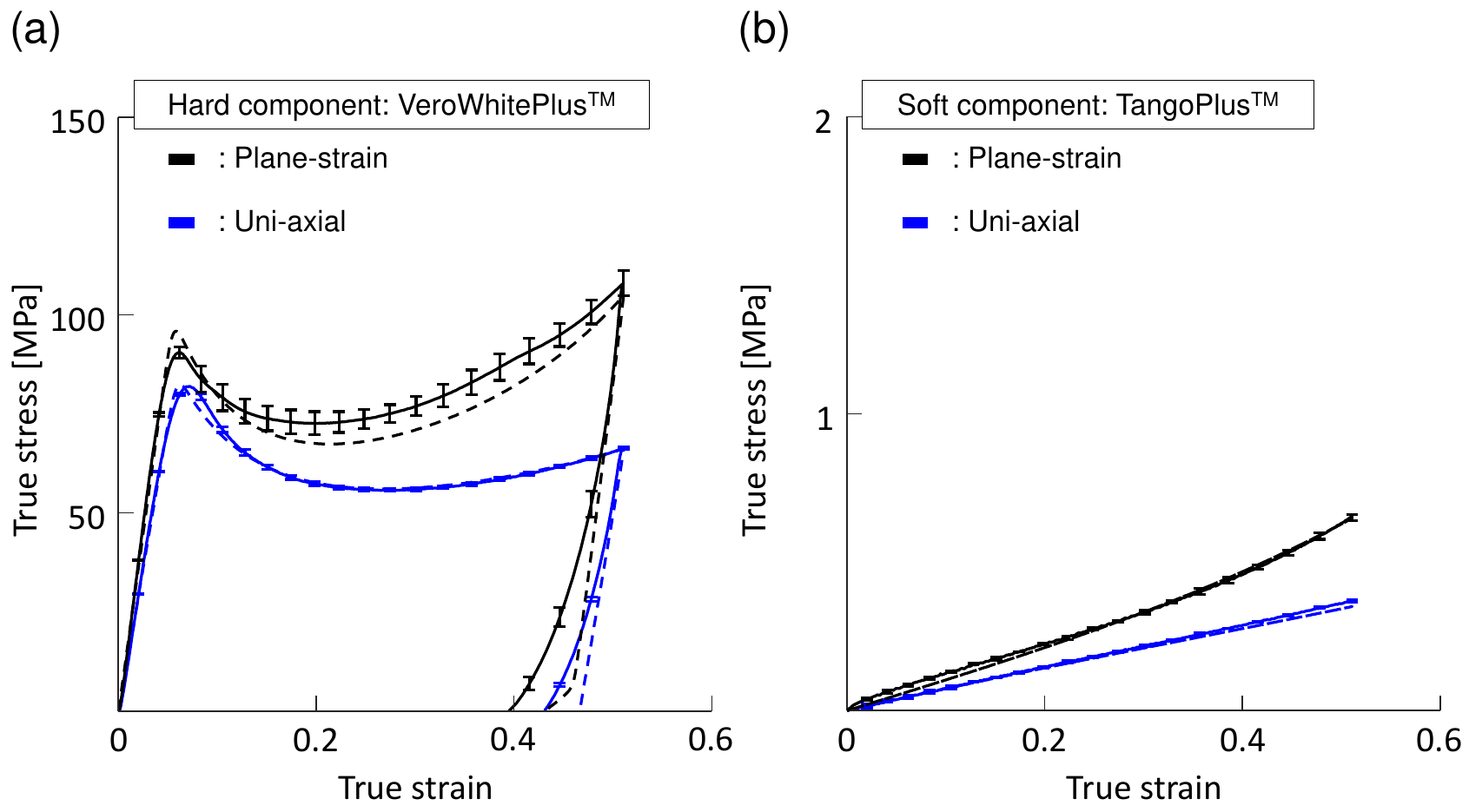}
    \caption{Stress-strain curves of (a) hard and (b) soft components under uniaxial (blue lines) and plane strain (black lines) compression at a strain rate of 0.01$\mathrm{s}^{-1}$ (solid lines: experiment, dashed lines: simulation). The hard component exhibits a standard thermoplastic behavior characterized by an initially stiff elastic response followed by a plastic yield, strain softening and entropic hardening at a large strain \cite{boyce1988large}. The soft component exhibits a hyperelastic behavior.}
    \label{fig:vw_tp}
\end{figure}

\noindent Here, the plastic part of the velocity gradient is $\mathbf{L}^{p}_\mathrm{I} = \mathbf{D}^{p}_\mathrm{I} + \mathbf{W}^{p}_\mathrm{I}$, where $\mathbf{D}^{p}_\mathrm{I}$ is the rate of plastic stretching and $\mathbf{W}^{p}_\mathrm{I}$ is the plastic spin. Furthermore, we make two important kinematical assumptions for plastic flow; the flow is incompressible (i.e., $\operatorname{det}\mathbf{F}^p_{\mathrm{I}}=1$) and irrotational (i.e., $\mathbf{W}^{p}_\mathrm{I}=0$). Thus, the rate of change in the plastic deformation gradient is given by,
\begin{equation}
\begin{aligned}
        \dot{\mathbf{F}}^p_\mathrm{I} = \mathbf{D}^p_\mathrm{I} \mathbf{F}^{p}_\mathrm{I}.
\end{aligned}
\end{equation}

\noindent The Cauchy stress ($\mathbf{T}_{\mathrm{I}}$) in the time-dependent elastic-viscoplastic mechanism I is expressed by,
\begin{equation}
\begin{aligned}
\mathbf{T}_{\mathrm{I}} = \frac{1}{J} \mathbf{R}^{e}_{\mathrm{I}} \mathbf{M}^{e}_{\mathrm{I}} \mathbf{R}^{e \top}_{\mathrm{I}},
\end{aligned}
\end{equation}

\noindent with the Mandel stress $\mathbf{M}^{e}_{\mathrm{I}} = 2\mu_{\mathrm{I}} (\ln \mathbf{U}^{e}_{\mathrm{I}})_{0} \, + \, K(\ln J)\mathbf{I}$, the elastic shear modulus $\mu_\mathrm{I}$ and the bulk modulus $K$. We note that the total bulk response is lumped into the mechanism I. The rate of plastic stretching $\mathbf{D}^p_\mathrm{I}$ is assumed to be coaxial to the deviatoric part of the Mandel stress (i.e., $(\mathbf{M}^e_\mathrm{I})_0 = \mathbf{M}^e_\mathrm{I} - \frac{1}{3}\mathrm{tr}(\mathbf{M}^e_\mathrm{I})\mathbf{I}$),

\begin{equation}
\mathbf{D}^{p}_{\mathrm{I}} =\frac{\dot{\gamma}^{p}_{\mathrm{I}}}{\sqrt{2}}\mathbf{N}^{p}_{\mathrm{I}}, \quad \text{where} \quad \mathbf{N}^{p}_{\mathrm{I}} = \frac{(\mathbf{M}^{e}_{\mathrm{I}})_{0}}{\|(\mathbf{M}^{e}_{\mathrm{I}})_{0}\|} \quad \text{and} \quad\|(\mathbf{M}^{e}_{\mathrm{I}})_{0}\| = \sqrt{(\mathbf{M}^{e}_{\mathrm{I}})_{0} : (\mathbf{M}^{e}_{\mathrm{I}})_{0}}.
\end{equation}
Then, we employ the thermally-activated viscoplasticity model prescribed by,
\begin{equation}
    \dot{\gamma}^{p}_{\mathrm{I}} = \dot{\gamma}_{\mathrm{I},0} \operatorname{exp}\left[ -\frac{\Delta G}{k\theta} \left\{1-\frac{\bar{\tau}}{s_{\mathrm{I}}}  \right\}   \right], \quad \text{where} \quad \bar{\tau} = \frac{1}{\sqrt{2}} \|(\mathbf{M}^{e}_{\mathrm{I}})_{0}\|,
\end{equation}
with the reference plastic strain rate $\dot{\gamma}_{\mathrm{I},0}$, the activation energy $\Delta G$, Boltzmann's constant $k$, the absolute temperature $\theta=295\mathrm{K}$ (room temperature), the shear strength $s_{\mathrm{I}}$ and the magnitude of the deviatoric part of the Mandel stress $\bar{\tau}$. Furthermore, to capture the strain softening beyond the initial yield in the hard component, we employ a simple saturation-type evolution rule for $s_{\mathrm{I}}$,
\begin{equation}
    \dot{s}_{\mathrm{I}} = h_{\mathrm{I}}(1-\frac{s_{\mathrm{I}}}{s_{\mathrm{I},ss}})\dot{\gamma}^{p}_{\mathrm{I}},
\end{equation}
with the softening slope $h_{\mathrm{I}}$ and the saturated shear strength $s_{\mathrm{I},ss}$. The initial value of shear strength is identified to be $s_{\mathrm{I},0}=66.8$MPa.

The (deviatoric) Cauchy stress ($\mathbf{T}_{\mathrm{N}}$) in the time-independent hyperelastic mechanism N is given by,
\begin{equation}
\mathbf{T}_{\mathrm{N}} = \frac{\mu_{\mathrm{N}}}{3J} \frac{\lambda_{\mathrm{N}}}{\bar{\lambda}} \mathscr{L}^{-1} \left(\frac{\bar{\lambda}}{\lambda_{\mathrm{N}}}\right) (\bar{\mathbf{B}}_{\mathrm{N}})_0 \quad \text{where} \quad \bar{\lambda} = \sqrt{\frac{\text{tr}\bar{\mathbf{B}}_{\mathrm{N}}}{3}} \quad \text{and} \quad (\bar{\mathbf{B}}_{\mathrm{N}})_0 = \bar{\mathbf{B}}_{\mathrm{N}} - \frac{1}{3}\mathrm{tr}(\bar{\mathbf{B}}_{\mathrm{N}})\mathbf{I},
\end{equation}
with the elastic shear modulus $\mu_{\mathrm{N}}$ and the limiting chain extensibility $\lambda_{\mathrm{N}}=\sqrt{N_{\mathrm{N}}}$; $\mathscr{L}^{-1}$ is the inverse Langevin function with $\mathscr{L}(x)=\coth(x)-\dfrac{1}{x}$.

\begin{table}[b!]
\centering
\small{
\begin{tabular}{lcc}
\hline
\textbf{Hard component: Time-dependent elastic-viscoplastic mechanism I} &  &  \\
\hline \\
$\mathbf{T}_{\mathrm{I}} = \frac{1}{J} \mathbf{R}^{e}_{\mathrm{I}} \mathbf{M}^{e}_{\mathrm{I}} \mathbf{R}^{e \top}_{\mathrm{I}}, \quad \text{where} \quad \mathbf{M}^{e}_{\mathrm{I}} = 2\mu_{\mathrm{I}} (\ln \mathbf{U}^{e}_{\mathrm{I}})_{0} \, + \, K(\ln J)\mathbf{I}$ & $\mu_{\mathrm{I}}$ [MPa] & 486.1 \\
& $K$ [GPa] & 3.97 \\
$\dot{\gamma}^{p}_{\mathrm{I}} = \dot{\gamma}_{\mathrm{I},0} \operatorname{exp}\left[ -\frac{\Delta G}{k\theta} \left\{1-\frac{\bar{\tau}}{s_{\mathrm{I}}}  \right\}   \right], \quad \text{where} \quad \bar{\tau} = \frac{1}{\sqrt{2}} \|(\mathbf{M}^{e}_{\mathrm{I}})_{0}\|$ & $\Delta G$ [$10^{-19}$J] & 1.2 \\
& $\dot{\gamma}_{\mathrm{I},0}$ [$10^{2}s^{-1}$] & 1.26 \\
$\dot{s}_{\mathrm{I}} = h_{\mathrm{I}}(1-\frac{s_{\mathrm{I}}}{s_{\mathrm{I},ss}})\dot{\gamma}^{p}_{\mathrm{I}}$ & $s_{\mathrm{I},ss}$ [MPa] & 0.45$s_{\mathrm{I},0}$ \\
& $h_{\mathrm{I}}$ [MPa] & 150 \\
\hline
\textbf{Hard component: Time-independent hyperelastic mechanism N} &  &   \\
\hline \\
$\mathbf{T}_{\mathrm{N}} = \frac{\mu_{\mathrm{N}}}{3J} \frac{\lambda_{\mathrm{N}}}{\bar{\lambda}} \mathscr{L}^{-1} \left(\frac{\bar{\lambda}}{\lambda_{\mathrm{N}}}\right) (\bar{\mathbf{B}}_{\mathrm{N}})_0$ & $\mu_{\mathrm{N}}$ [MPa] & 10.4  \\
& $\lambda_{\mathrm{N}}=\sqrt{N_\mathrm{\mathrm{N}}}$ & $\sqrt{2}$ \\
\hline
\textbf{Soft component} &  &   \\
\hline
& $\mu_{\mathrm{S}}$ [MPa] & 0.235  \\
$ \mathbf{T}_{\mathrm{S}}=\frac{\mu_\mathrm{S}}{3 J} \frac{\lambda_\mathrm{S}}{\bar{\lambda}} \mathscr{L}^{-1}\left(\frac{\bar{\lambda}}{\lambda_\mathrm{S}}\right)\left(\bar{\mathbf{B}}\right)_0 + {K_\mathrm{S}} (J-1) \mathbf{I}$ & $K_\mathrm{S}$ [MPa] & 16.71  \\
& $\lambda_\mathrm{S}=\sqrt{N_\mathrm{S}}$ & $\sqrt{6.5}$ \\
\hline
\end{tabular}
}
\caption{Material parameters used in the constitutive models for hard and soft components.}
\label{Tab:material_parameter}
\end{table}

The total stress in the hard component is then obtained by,
\begin{equation}
\mathbf{T}_{\mathrm{hard}} = \mathbf{T}_{\mathrm{I}} + \mathbf{T}_{\mathrm{N}}.
\end{equation}

A nearly incompressible Arruda-Boyce model \cite{arruda1993three} is used for the soft component. The Cauchy stress in the soft component is expressed by,
\begin{equation}
\begin{aligned}
    \mathbf{T}_{\mathrm{S}}=\frac{\mu_\mathrm{S}}{3 J} \frac{\lambda_\mathrm{S}}{\bar{\lambda}} \mathscr{L}^{-1}\left(\frac{\bar{\lambda}}{\lambda_\mathrm{S}}\right)\left(\bar{\mathbf{B}}\right)_0 + {K_\mathrm{S}} (J-1) \mathbf{I} \quad &\text{where} \quad \bar{\lambda} = \sqrt{\frac{\mathrm{tr}(\bar{\mathbf{B}})}{3}},
\end{aligned}
\end{equation}
where $\mu_\mathrm{S}$ is the elastic shear modulus, $K_\mathrm{S}$ is the bulk modulus, $\lambda_\mathrm{S}=\sqrt{N_\mathrm{S}}$ is the limiting chain extensibility where $N_\mathrm{S}$ is the average number of effective chain segments, $\bar{\mathbf{B}} = J^{-2/3} \mathbf{F} \mathbf{F}^{\top}$, $\mathbf{F}$ is the deformation gradient and $\bar{\mathbf{B}}_0$ is the deviatoric part of the isochoric left Cauchy-Green tensor (i.e., $\bar{\mathbf{B}}_0 = \bar{\mathbf{B}} - \frac{1}{3}\mathrm{tr}(\bar{\mathbf{B}})\mathbf{I}$) in the soft component.

Figure \ref{fig:vw_tp} presents the stress-strain curves of both hard and soft components under uniaxial (blue lines) and plane strain (black lines) compression conditions obtained from experiments (solid lines) and numerical simulations (dashed lines). The material parameters used in the constitutive models are given in Table \ref{Tab:material_parameter}. Detailed information on the time integration procedures for the time-dependent elastic-viscoplastic mechanism I especially in the hard component can be found in \cite{weber1990finite}.
\clearpage

\section{Further analysis on energy dissipation, local failure and reusability}
\label{Further analysis on energy dissipation, local failure and reusability}
\renewcommand{\thefigure}{D.\arabic{figure}}
\renewcommand{\thetable}{D.\arabic{table}}
\renewcommand{\theequation}{D.\arabic{equation}}
\renewcommand{\thesubsection}{D.\arabic{subsection}}
\subsection{Reusability (cyclic behavior) with no idling time for shape recovery}
\label{section:reusability}
\begin{figure}[b!]
    \centering
    \includegraphics[width=1.0\textwidth]{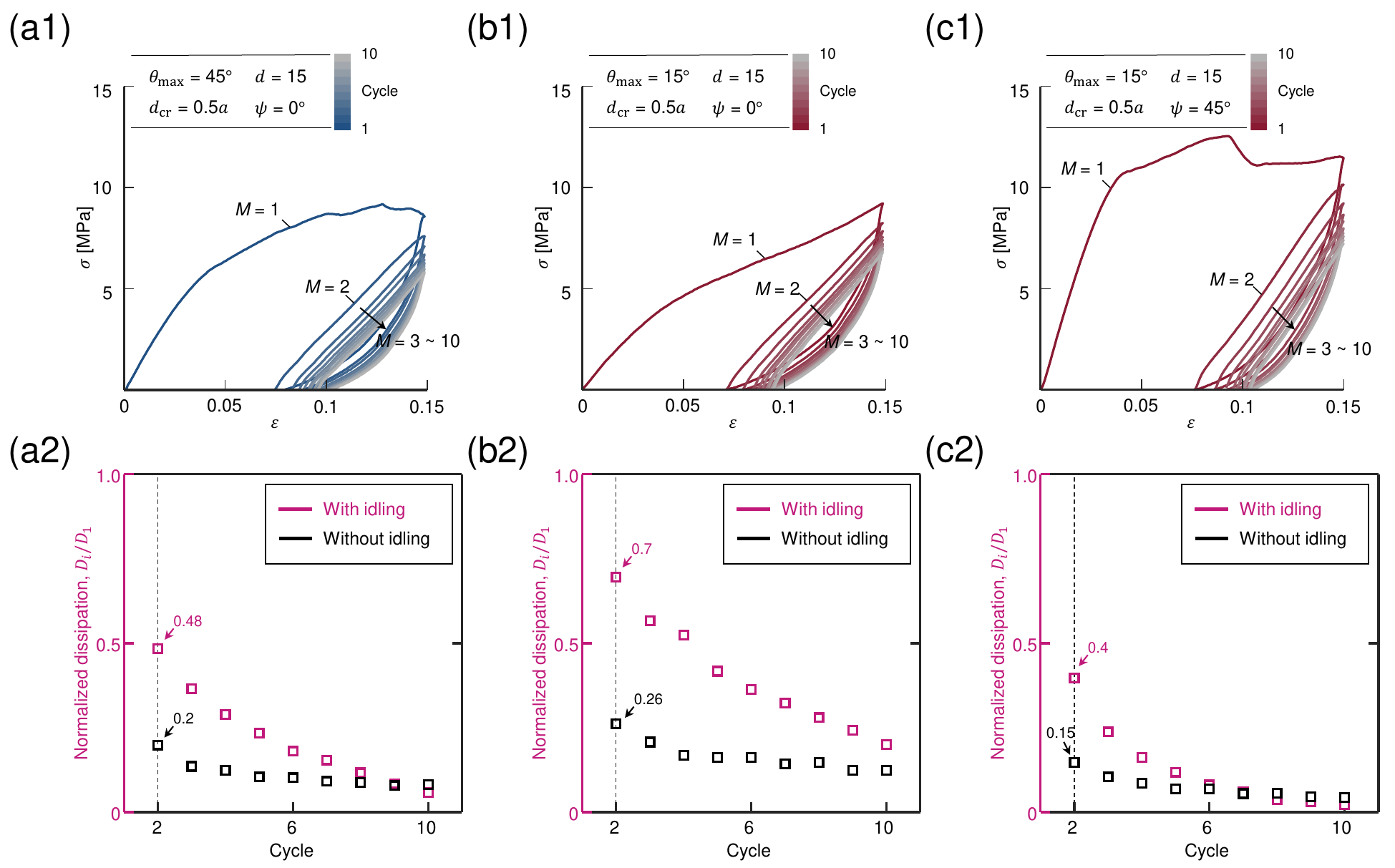}
    \caption{Reusability of heterogeneous architected polycrystals with no idling period. Stress-strain curves during multiple loading and unloading cycles of $M$ = 1 to 10 at a strain rate of 0.01$\mathrm{s}^{-1}$ together with normalized energy dissipation relative to that when $M$ = 1 in (a1) and (a2) weakly textured ($\theta_{\mathrm{max}}=45^\circ$) polycrystalline RVEs loaded at $\psi=0^\circ$ and strongly textured ($\theta_{\mathrm{max}}=15^\circ$) RVEs loaded at (b1) and (b2) $\psi=0^\circ$ and (c1) and (c2) $\psi=45^\circ$; here, the grain size and the critical distance are set to $d=15$ and $d_{\mathrm{cr}}=0.5a$.}
    \label{fig:cyclic}
\end{figure}
The reusability and energy dissipation capabilities of heterogeneous architected polycrystals presented in Figures \ref{fig:reusab_low}, \ref{fig:reusab_high_0deg} and \ref{fig:reusab_high_45deg} (i.e., $d=15$ and $d_{\mathrm{cr}}=0.5a$) were further investigated through cyclic loading-unloading mechanical testing ($M = 1 \sim 10$) with no idling period for inelastic shape recovery between cycles. The stress-strain curves in the weakly textured ($\theta_{\mathrm{max}}=45^\circ$) polycrystalline RVEs loaded at $\psi=0^\circ$ (Figure \ref{fig:cyclic}a1) and the strongly textured ($\theta_{\mathrm{max}}=15^\circ$) RVEs loaded at $\psi=0^\circ$ (Figure \ref{fig:cyclic}b1) and $\psi=45^\circ$ (Figure \ref{fig:cyclic}c1) are shown. The reloading curves ($M = 2 \sim 10$) in all of these polycrystalline RVEs exhibit significant degradation in the load transfer capability under these harsh, repeated loading and unloading cycles. Furthermore, the stress-strain data are further reduced to the normalized energy dissipation densities, relative to those when $M = 1$, as plotted in Figures \ref{fig:cyclic}a2, \ref{fig:cyclic}b2 and \ref{fig:cyclic}c2. All the polycrystalline RVEs with the idling time for shape recovery displayed in Figures \ref{fig:reusab_low}, \ref{fig:reusab_high_0deg} and \ref{fig:reusab_high_45deg} in the main text are found to exhibit much better reusability and energy dissipation capability compared to without any shape recovery between cycles.

\subsection{Stress concentration at grain boundaries with high $\Delta \varphi$}
\label{Stress concentration at grain boundaries}
\label{section:stress}
\begin{figure}[b!]
    \centering
    \includegraphics[width=1.0\textwidth]{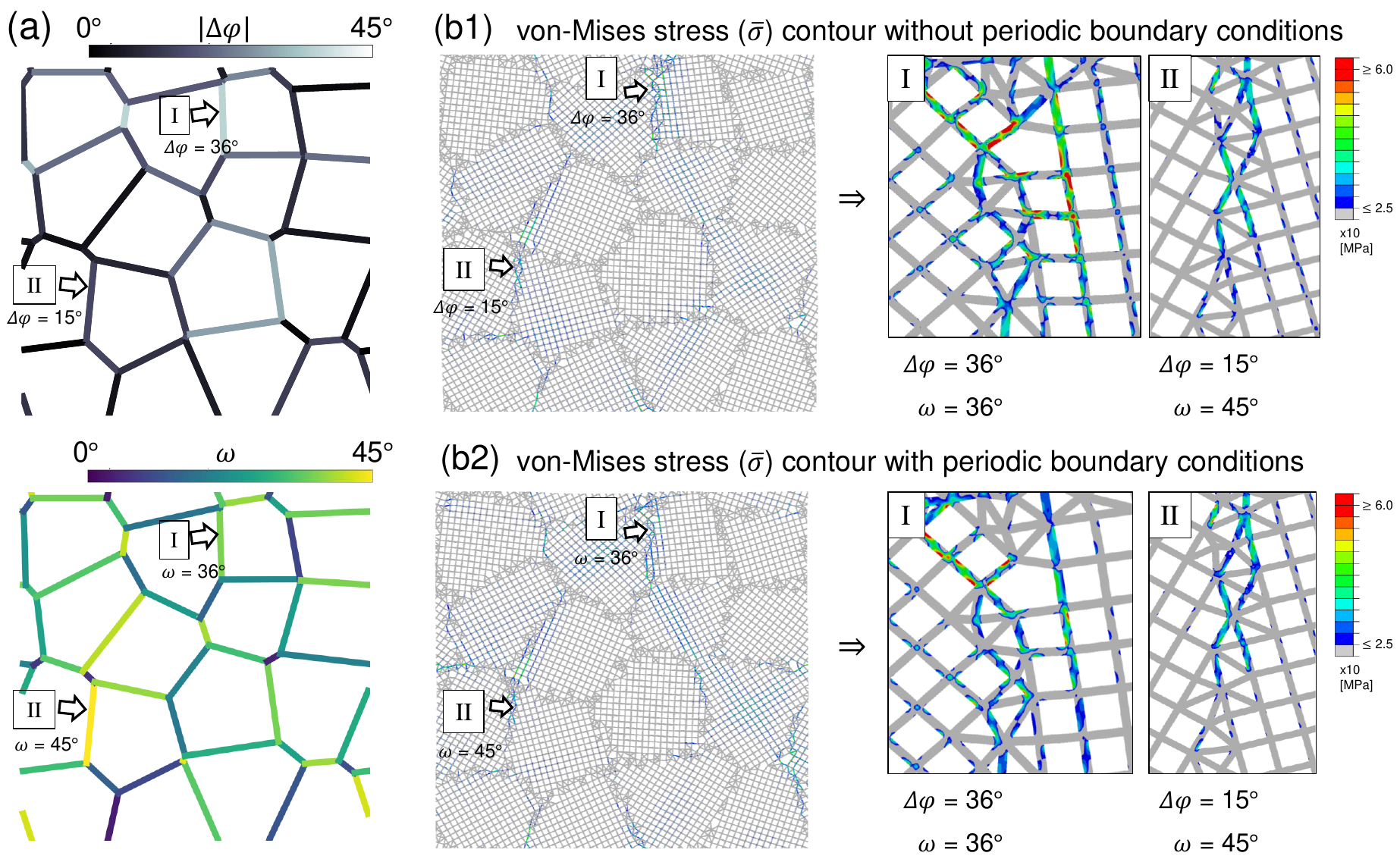}
    \caption{Stress concentration at grain boundaries with high $|\Delta \varphi|$ in weakly textured ($\theta_{\mathrm{max}}=45^\circ$) polycrystalline RVEs with $d=15$ and $d_{\mathrm{cr}}=0.5a$ loaded at $\psi=0^\circ$. (a) Distributions of $|\Delta \varphi|$ (upper) and $\omega$ (lower) in neighboring grain pairs. The von Mises stress contours from numerical simulations conducted (b1) without and (b2) with periodic boundary conditions at a macroscopic strain of 0.02; here, only hard domains are displayed.}
    \label{fig:stress_concentration}
\end{figure}
We here present the von Mises stress contours throughout the hard domains (i.e., polycrystalline networks) in the weakly textured ($\theta_{\mathrm{max}}=45^\circ$) polycrystalline RVEs with $d=15$ and $d_{\mathrm{cr}}=0.5a$ loaded at $\psi=0^\circ$; the corresponding results are presented in Figure \ref{fig:reusab_low} in the main text. In particular, stress distributions at two representative grain boundaries are examined: (i) a grain boundary region with a high $|\Delta \varphi|$ (i.e., $|\Delta \varphi|=36^\circ$) and (ii) a grain boundary region with a high $\omega$ (i.e., $\omega=45^\circ$), denoted I and II in Figure \ref{fig:stress_concentration}a on the distributions of $|\Delta \varphi|$ (upper inset) and $\omega$ (lower inset). Figures \ref{fig:stress_concentration}b1 and \ref{fig:stress_concentration}b2 show the von Mises stress contours at a macroscopic strain of 0.02, taken from numerical simulations conducted without and with periodic boundary conditions. Significantly larger stress concentrations are observed at the grain boundary with $|\Delta \varphi|=36^\circ$ (I) than at the grain boundary with $\omega=45^\circ$ (II), further supporting that the local failure is initiated at the grain boundary with the high $|\Delta \varphi|$ between neighboring grains, presented in Figure \ref{fig:reusab_low}.
\clearpage

\section{List of supplementary movies}
\label{List of supplementary movies}
All movies are available online in the following repository. \\
\href{https://solidslabkaist.github.io/archives/movies/architected_polycrystal}{https://solidslabkaist.github.io/archives/movies/architected\_polycrystal}
\\

\noindent
$\mathbf{Movie~S1.}$ Constructing the architected polycrystals by mimicking the crystal growth. The atoms are radially arranged, and removed or merged in the grain boundary regions within a Voronoi-tessellated $N = 2 \times 2$ polycrystal. See Figure \ref{fig:design} and Figure \ref{fig:gb_strength} for specific details on the merging procedure of the neighboring atoms in the grain boundary regions.
\vspace*{0.3in}

\noindent
$\mathbf{Movie~S2.}$ Weakly textured ($\theta_{\mathrm{max}}=45^\circ$) heterogeneous architected polycrystal with $d=15$ and $d_{\mathrm{cr}}=0.5a$ loaded at $\psi=0^\circ$. The experimental images and the von Mises strain fields show that the local failures, denoted I, II and III, initiate at the “grain boundaries” with significant inter-grain inhomogeneity (i.e., high $|\Delta \varphi|$ between the neighboring grains). Furthermore, the local failure propagation stops at the high-angle grain boundaries. See Figure \ref{fig:reusab_low} for more details.
\vspace*{0.01in}

\noindent
$\mathbf{Movie~S3.}$ Strongly textured ($\theta_{\mathrm{max}}=15^\circ$) architected polycrystal with $d=15$ and $d_{\mathrm{cr}}=0.5a$ loaded at $\psi=0^\circ$. It exhibits relatively high reusability since there are no significant local failures throughout the polycrystalline network, as supported by the relatively uniform strain distribution without any significant localization. See Figure \ref{fig:reusab_high_0deg} for more details.
\vspace*{0.3in}

\noindent
$\mathbf{Movie~S4.}$ Strongly textured ($\theta_{\mathrm{max}}=15^\circ$) architected polycrystal with $d=15$ and $d_{\mathrm{cr}}=0.5a$ loaded at $\psi=45^\circ$. As shown in Figures \ref{fig:reusab_high_45deg}a1 and \ref{fig:reusab_high_45deg}a2, the load transfer and energy dissipation capabilities significantly decrease between the first two cycles. This is further supported by the local failures, denoted I and II, which mainly initiate in the “grain interiors”, as shown in the DIC images. The local failures propagate rapidly throughout the neighboring grains in this architected polycrystal with strong crystallographic texture. See Figure \ref{fig:reusab_high_45deg} for more details.
\clearpage
\bibliographystyle{elsarticle-num}
\bibliography{refs}
\end{document}